\documentclass[sn-nature]{sn-jnl}

\usepackage{graphicx}
\usepackage{multirow}
\usepackage{amsmath,amssymb,amsfonts}
\usepackage{amsthm}
\usepackage{mathrsfs}
\usepackage[title]{appendix}
\usepackage{xcolor}
\usepackage{textcomp}
\usepackage{manyfoot}
\usepackage{booktabs}
\usepackage{algorithm}
\usepackage{algorithmicx}
\usepackage{algpseudocode}
\usepackage{listings}
\usepackage{xspace}
\usepackage{chemformula}
\usepackage{caption}
\usepackage{subcaption}
\usepackage{adjustbox}
\usepackage{rotating}
\usepackage{tabularx}
\usepackage{pdflscape}
\usepackage{gensymb}
\usepackage{tikz}
\usepackage{upgreek}
\usepackage[normalem]{ulem}
\usepackage{nameref}
\usepackage{booktabs}
\usepackage[table]{xcolor}

\def\aj{AJ}

\def\apj{ApJ}
\def\apjl{ApJ}

\def\aap{A\&A}

\def\icarus{Icarus}

\def\mnras{MNRAS}

\def\pasp{PASP}

\def\ssr{Space~Sci.~Rev.}

\def\nat{Nature}

\newcommand{\pds}{PDS 70\xspace}

\makeatletter
\providecommand{\allowdisplaybreaks}[1][]{}
\makeatother

\begin{document}

\title[Article Title]{Potential sublimating exocomets around the young star PDS 70}

\author*[1]{\fnm{Aline} \sur{Novais}}\email{aline.novais@fysik.lu.se}

\author[1]{H. \fnm{Jens} \sur{Hoeijmakers}}

\author[1]{\fnm{Alexandra} Stockwell \sur{Murphy}}

\author[1,2]{\fnm{Bibiana} \sur{Prinoth}}

\author[1,3]{\fnm{Anders} \sur{Johansen}}

\author[1]{\fnm{Klaudia} \sur{Jaworska}}

\author[4]{\fnm{Luan} \sur{Ghezzi}}

\affil[1]{\orgdiv{Division of Astrophysics, Department of Physics}, \orgname{Lund University}, \orgaddress{\street{Box 118}, \postcode{221 00}, \city{Lund}, \country{Sweden}}}

\affil[2]{\orgname{European Southern Observatory}, \orgaddress{\street{Karl-Schwarzschild-Strasse 2}, \postcode{85748}, \city{Garching}, \country{Germany}}}

\affil[3]{\orgdiv{Center for Star and Planet Formation, Globe Institute}, \orgname{University of Copenhagen}, \orgaddress{\street{Øster Voldgade 5–7}, \postcode{1350}, \city{Copenhagen}, \country{Denmark}}}

\affil[4]{\orgdiv{Valongo Observatory}, \orgname{Federal University of Rio de Janeiro}, \orgaddress{\street{Ladeira do Pedro Antonio, 43}, \postcode{20080-090}, \city{Rio de Janeiro}, \country{Brazil}}}

\abstract{
Recent observations by the James Webb Space Telescope (JWST) have indicated the presence of water in the inner regions of the disc around the young protostar PDS 70, but the origin of this water remains unclear. Here we report the discovery of variable absorption lines of neutral sodium in archival High Accuracy Radial velocity Planet Searcher (HARPS) spectra of PDS 70. These lines vary strongly and stochastically on a daily basis, in amplitude, number, and radial velocity. Our measurements indicate that this gas tends to be optically thick, and partially cover the stellar disc, meaning this fast-moving gas is often spatially confined. We explore several hypotheses for the origin of these lines, and conclude that a likely source of the observed sodium is sublimation of planetesimals that transit the star on highly elliptical orbits, reminiscent of the exocomet phenomenon seen in other seen in other extrasolar systems. These sublimating exocomets could play a role in sourcing the previously observed water in the inner, terrestrial-planet forming region of the PDS 70 system.
}

\maketitle

\section*{Introduction}

PDS 70 is a 5.4 Myr K7 T Tauri star located in the direction of the Centaurus constellation, with a measured mass and radius of $0.76 \pm 0.02$ M$_\odot$ \citep{Muller+18} and $1.26 \pm 0.15$ R$_\odot$ \citep{Keppler+18}. As is common for young stellar objects, PDS 70 holds a transition disc (i.e. a protoplanetary disc with gaps in the dust distribution), which may be an effect of the ongoing formation of planets in the system \citep{Keppler+18, Muller+18}. Indeed, two giant-mass protoplanets PDS 70 b and PDS 70 c have been detected orbiting the protostar at semi-major axes of $20.6 \pm 1.2$ au and $34.5 \pm 2.0$ au, respectively \citep{Haffert+19}. These planets are both located inside a 36 au-wide gap, which separates the inner disc -- that extends up to 18 au from the protostar -- from the outer disc -- that spans from around 54 to 87 au away from the T Tauri object \citep{Benisty+21, Perotti+23}. A third protoplanet was proposed to exist inside the inner disc at an orbital distance of $13.5^{+0.3}_{-0.2}$ au \citep{Mesa+19, Christiaens+24}, roughly placing it at a 1:2:4 mean-motion resonance configuration with PDS 70 b and c \citep{Wang+21}, yet its existence is presently unconfirmed.\\

In the very centre of the inner disc of PDS 70, recent infrared observations using the Mid-InfraRed Instrument (MIRI) aboard the James Webb Space Telescope (JWST) revealed spectroscopic emission signatures of a gaseous water reservoir \citep{Perotti+23}. With an estimated temperature of 600 K, this water is inferred to extend to approximately 0.05 au from the star, lying interior to a region where potential terrestrial planets could be formed (Fig. \ref{fig1}a). Theories about where this water originates include in-situ formation in the inner disc via chemical reactions between \ch{O}, \ch{H2}, and/or \ch{OH}, as well as delivery via icy dust particles that are dragged along with the gas from the outer disc, across the gap opened by PDS 70 b and c, and towards the inner disc \citep{Perotti+23, Pinilla+24}. However, an alternative, yet unexplored possibility is that a fraction of this water was transported from outer regions by planetesimals, which are scattered towards the inner disc after dynamical interactions with the planets.\\

Motivated by the transport hypothesis, our study investigates if planetesimals in PDS 70 could be a source of delivery of water and volatiles to the stellar vicinity by searching for exocomet activity in stellar observations. Exocomets are comet-like planetesimals in eccentric orbits that sublimate material when in the vicinity of their star \citep{Iglesias+25}, mainly observed in transit in extrasolar systems. These have been extensively studied via spectroscopic observations \citep{Ferlet+87, LagrangeHenri+89, VidalMadjar+94, MontgomeryWelsh12, Kiefer+14, Kiefer+14b, WelshMontgomery16, Strom+20, Rebollido+20, Vrignaud+24, VrignaudLecavelierdesEtangs24, Vrignaud+25, Iglesias+25}, and more recently in an increasing number of systems through photometric monitoring \citep{Zieba+19, Pavlenko+22, LecavelierdesEtangs+22, Kiefer+23, Rebollido+23, Dumond+25, Gibson+25}. In spectroscopy, exocomets are characterised by time-sensitive radial velocity variations of absorption lines of gas-phase metals \citep{WelshMontgomery16, Vrignaud+24, VrignaudLecavelierdesEtangs24, Vrignaud+25}, which notably appear and disappear on a day-to-day timescale \citep[e.g.][]{Ferlet+87, LagrangeHenri+89, VidalMadjar+94, MontgomeryWelsh12, Kiefer+14, Kiefer+14b}. These lines are also strongly Doppler-shifted, indicating that these exocomets transit the star at high velocities on eccentric orbits. Exocomet activity has so far been observed in various absorption lines of neutral and ionised metals \citep{WelshMontgomery16, Vrignaud+24, VrignaudLecavelierdesEtangs24, Vrignaud+25} including sodium \citep{Hoeijmakers+25}, in particular in the benchmark system $\beta$ Pictoris, where exocomet activity is presently ongoing vigorously.\\

In this work we examine the exocomet hypothesis in the PDS 70 system via the analysis of transient Na\,I absorption lines in HARPS data obtained in 2018 (see subsection \nameref{sec:properties_variable_absorption_lines}). We measure the properties of these variable absorption lines of sodium at each epoch of 2018, and use these properties to investigate the origin of these lines by comparing them with disc wind absorption lines previously reported in PDS 70. We additionally compare PDS 70 with other T Tauri stars that present variability, and conclude the variable Na\,I absorption lines of PDS 70 are likely exocometary. We proceed on further analysing the exocomet scenario by investigating the characteristics of the PDS 70 variable Na\,I lines in comparison with the well-studied exocomet system $\beta$ Pictoris. Finally, we perform N-body simulations to evaluate if exocomets in PDS 70 could be scattered by the planets towards the inner disc, in line with the hypothesis of water and volatile delivery to the inner regions of the system.\\

\section*{Results and discussion}

\subsection*{Properties of the variable sodium absorption lines}
\label{sec:properties_variable_absorption_lines}

We revisit high-resolution archival data of PDS 70 in search for the exocomet phenomenon. We examine data taken with the High Accuracy Radial velocity Planet Searcher (HARPS) spectrograph on the European Southern Observatory's (ESO) 3.6 m telescope mounted on La Silla, Chile \citep{Mayor+03}, covering a wavelength range between 380 and 690 nm with a resolving power of $R = 115\,000$. A total of 52 spectra were taken on 22 different nights in the first months of 2018, 2019, and 2020 (Programme ID 098.C-0739, 101.C-0557, 1101.C-0557, 0104.C-0418, PI A.-M. Lagrange), with most systematic coverage in 2018 (12 nights) and 2020 (8 nights) compared to 2019 (2 nights). The 2018 data span a period of approximately 46 days, while the 2020 data span a period of 20 days. All 52 spectra are displayed in Fig. \ref{fig2}. A detailed observation log is listed in Table \ref{table1}, while information about data treatment is described in Methods (see subsection \nameref{sec:observations_analysis_methods}).\\

A detailed inspection of the archival HARPS data across the full spectral range reveals transient absorption signals that vary strongly in number, strength, and radial velocity in the region around the neutral sodium (Na\,I) Fraunhofer D lines, centred at 588.996 and 589.593 nm, over the 12 nights of observation in 2018 (Fig. \ref{fig1}b and Fig. \ref{fig2}). Each epoch in 2018 displays a stochastic appearance and disappearance of blue-shifted Na\,I absorption lines embedded in the broad Na\,I absorption line doublet of the stellar photosphere, resembling absorption lines associated with exocomet activity \citep{Ferlet+87, LagrangeHenri+89, VidalMadjar+94, MontgomeryWelsh12, Kiefer+14, Kiefer+14b, WelshMontgomery16, Strom+20, Rebollido+20, Vrignaud+24, VrignaudLecavelierdesEtangs24, Vrignaud+25, Iglesias+25, Hoeijmakers+25}. This stochastic Doppler-shifted variability is not seen in any other species covered by HARPS. The two nights of data in 2019 similarly show blue-shifted Na\,I absorption lines, but with only two nights of observations, variability at these epochs cannot be confidently established. In 2020, this variable absorption has largely disappeared. New high-resolution observations obtained in 2026 with the Ultraviolet and Visual Echelle Spectrograph (UVES, Programme ID 0116.C-0329, PI A. Novais) indicate that this activity continues intensively at the present time (Fig. \ref{fig3}). While the analysis in this study is primarily aimed at ascertaining the exocomet interpretation in the 2018 HARPS spectra of PDS 70, follow-up studies will focus on a detailed analysis and discussion of the UVES dataset.\\

PDS 70 exhibits a broad photospheric Na\,I absorption line across all observed epochs in 2018, 2019, and 2020. Superimposed in the photospheric line, all HARPS spectra present an emission line approximately at the stellar rest frame, and two deep, narrow, partially blended absorption lines blue-shifted with respect to the stellar rest frame. We refer to these as stationary emission and stationary absorption lines throughout the text (see Methods, subsection \nameref{sec:stationary_lines_methods} for a detailed analysis). The HARPS data obtained in 2018, spanning 1.5 months of sporadic monitoring, reveal additional absorption lines of Na\,I embedded in the photospheric absorption line of PDS 70 (Fig. \ref{fig4} and Fig. \ref{fig5}), Doppler-shifted to radial velocities that vary on a daily basis, meaning that spectra taken in consecutive days are notably dissimilar. This day-to-day variability in radial velocity is easily distinguishable from the stationary absorption lines, which are observed to be blue-shifted at the same radial velocity at all epochs, including in the absence of variable absorption lines in 2020 (Fig. \ref{fig2}). We refer to the additional, stochastically variable absorption lines present in the 2018 data as variable absorption lines.\\

We constructed a model to fit the properties of all stationary and variable Na\,I line components present in each spectrum, in particular their radial velocity, width, and optical depth at line centre (see Methods, subsections \nameref{sec:observations_analysis_methods} and \nameref{sec:fitting_methods}). Many of the absorption lines superimposed on the photospheric Na\,I line in the HARPS data appear to have high optical depths, resulting in partial saturation. The Na\,I doublet lines have unequal intrinsic oscillator strengths with a ratio of approximately 2 \citep{Forbrich+68}. Assuming the absorbing gas to be homogeneous, if the absorption is optically thin, this results in a line ratio that is approximately equal to 2. In the optically thick regime, both lines saturate, and the line ratio approaches unity. In this case, if the gas producing these absorption lines does not entirely cover the stellar disc, then the flux at the centre of these saturated lines is greater than zero, making the line depths sensitive to the surface fill fraction, i.e. the fraction of the stellar disc being obscured by the cloud of gas, in the optically thick regime. We include the optical depth at line centre and surface fill fraction as free parameters in our model \citep{LagrangeHenri+89, Kiefer+14, Strom+20}, which means they are expected to be degenerate \citep{Hoeijmakers+25}.\\

Fig. \ref{fig4} shows an example of nine epochs of PDS 70 observed within 50 hours in March 29, 30, and 31, 2018, where at least two variable Na\,I absorption lines are present in each epoch. In the very first epoch (2018-03-29 (1)), the blue-most line component is Doppler-shifted to a velocity of $-80.93^{+0.23}_{-0.28}$ km/s. One hour later (2018-03-29 (3)), the same component shows a small deceleration to $-80.78^{+0.14}_{-0.18}$ km/s. 23 hours after the first epoch (2018-03-30 (1)), the red-most component visible on March 29 has disappeared, while the blue-most component in this epoch is visible at $-60.81^{+0.26}_{-0.23}$ km/s, which is a difference of $20.12^{+0.35}_{-0.36}$ km/s from the most blue-shifted component in the first epoch. After three hours (2018-03-30 (3)), the same component has decelerated to $-57.32^{+0.29}_{-0.30}$ km/s. Similarly, 45 hours after the first epoch (2018-03-31 (1)), the blue-most component is at $-57.69^{+0.02}_{-0.04}$ km/s, which is a difference of $23.24^{+0.23}_{-0.29}$ km/s from the very first epoch. However, after five hours (2018-03-31 (3)), the same component has accelerated to $-61.44^{+0.16}_{-0.15}$ km/s. These spectra reveal a strong night-to-night variability, including radial velocity shifts within the same night, both in acceleration and deceleration with respect to the star.\\

In a total of 18 epochs in 2018 (see Methods, subsection \nameref{sec:observations_analysis_methods} for number of epochs), we identify 43 Na\,I absorption line components that vary in Doppler-shift compared to the spectrum from the preceding epoch. These have radial velocities ranging from $-25$ to $-115$ km/s. It is possible that low-velocity absorption lines blue-shifted to up to $-25$ km/s are additionally present in these spectra, yet these would be hard to distinguish due to overlaps with the two stationary absorption lines at $-22$ and $-16$ km/s. A fit to the depths and ratios of these variable Na\,I absorption components reveals that most of these obscuring Na\,I gas clouds are optically thick (i.e. optical depth above unity), with surface fill fractions tightly constrained and significantly below unity (Fig. \ref{fig4}b and Fig. \ref{fig5}b). This allows us to establish that these sodium clouds fill the stellar disc by 10 to 90\%, with over 90\% of these measurements confidently below 80\%. The width of these line components range between 2 and 10 km/s (Fig. \ref{fig6}a). From the optical depths and line widths measured from of our variable Na\,I absorption lines, we estimate that the observed sodium clouds have column densities (i.e. the number of sodium atoms along the line of sight) between $2.4 \times 10^{11}$ and $7.1 \times 10^{12}$ cm$^{-2}$ (see Methods, subsection \nameref{sec:column_density_methods}). Assuming a homogeneous cloud of sodium gas in front of a spherical star, our measured column densities and surface fill fractions suggest these clouds have minimum masses between $1.3 \times 10^{8}$ and $3.9 \times 10^{9}$ kg of sodium (Fig. \ref{fig6}b, see Methods, subsection \nameref{sec:cloud_mass_methods}).\\

The rapid time variability in Doppler-shifted radial velocities observed in the variable Na\,I absorption lines of PDS 70 are consistent with previous detections of exocomets in other systems \citep[e.g.][]{Ferlet+87, LagrangeHenri+89, VidalMadjar+94, MontgomeryWelsh12, Kiefer+14, Kiefer+14b, Hoeijmakers+25}. A transiting material on a circular Keplerian orbit would move together with the disc, presenting a radial velocity equal to the stellar rest frame. A systematic Doppler-shifted absorption suggests that the transiting material has a velocity component along the line of sight, which requires a highly elliptical orbit. In addition, the gas sublimated from transiting exocomets is typically observed to be optically thick and to cover significant fractions of the stellar disc, but often without fully obscuring it \citep{LagrangeHenri+89, Vrignaud+24, Hoeijmakers+25}, indicating that the material is clumpy, partially covering the stellar disc. This is in accordance with the predominance of optical depths above unity and surface fill fractions considerably less than unity measured from the variable Na\,I absorption lines in HARPS spectra of PDS 70.\\

\subsection*{Evaluating the possibility of Na\,I disc winds}
\label{sec:discwinds}

A crucial aspect of the variable Na\,I absorption lines embedded in the photospheric Na\,I doublet of PDS 70 is that they are exclusively blue-shifted at all epochs of 2018, meaning that these absorbing gas clouds in front of the star are always moving towards the observer. Outflows from the disc are known to cause blue-shifted absorption in various absorption lines of T Tauri stars, including in the sodium doublet \citep{NattaGiovanardi90}, which could be challenging to confidently distinguish from potential exocomet lines. Thus, at first glance, the absence of red-shifted lines in 2018 could challenge the interpretation that these clouds are associated with transiting, sublimating material, and instead favour the interpretation that these are outflows of material from the circumstellar disc, referred to as disc winds.\\

Indeed, the presence of a disc wind has been previously inferred in the spectrum of PDS 70 from a blue-shifted absorption He\,I line at 1083 nm, which likely traces the innermost, hottest regions of the circumstellar disc. This wind may be caused by accretion or be magnetically-driven \citep{Thanathibodee+20, CampbellWhite+23}, and has been observed to have a radial velocity that varies from $-85$ km/s in April 26, 2019 \citep{Thanathibodee+20}, to $-277$ km/s in December 24, 2020, and to $-94$ km/s in February 7, 2021 \citep{CampbellWhite+23}. These three epochs are 8 months and 1.5 months apart, but there is currently not enough data to confidently establish that this variation also exists on a daily basis, as seen in our variable absorption lines of Na\,I. In addition, the observed disc winds may be driven by different mechanisms at different times, which would inherently result in different velocities \citep{Thanathibodee+20, CampbellWhite+23}. Winds may originate at different radii in the disc, depending on the species available in that region and on the wind-driving mechanism \citep{KwanFischer11}, meaning that winds that are visible in H$\alpha$ may be sourced from regions at radii greater than the source regions of helium \citep{KwanFischer11}. However, the H$\alpha$ absorption lines in PDS 70 present no signs of radial velocity variability, suggesting that the He\,I outflow originates in regions closer to the star.\\

To evaluate if the blue-shifted absorption lines of Na\,I observed in the HARPS spectra of PDS 70 could originate from the same mechanism as He\,I, we apply the same disc wind model used for He\,I \citep{Thanathibodee+20, Calvet97} to estimate the column densities expected for sodium disc wind lines in PDS 70. The magnitude of the mass loss rate of any disc wind can be estimated as being a fraction of the mass accretion rate of the star \citep{Calvet97, Edwards+06}, which is typically $\dot{M}_{\rm wind} / \dot{M}_{\rm acc} \simeq 0.1$--$0.3$ for moderate to low accretors such as PDS 70 \citep{NattaGiovanardi90, Calvet97, Rigliaco+13, IguchiItoh16}. Assuming this fraction is 0.1 for PDS 70, and assuming the minimum and maximum mass accretion rate measurements of $6.0 \times 10^{-11}$ and $2.2 \times 10^{-10}$ M$_\odot$/yr \citep{Thanathibodee+20}, we estimate the column density expected for sodium winds from the properties measured on each variable sodium absorption line in the 2018 spectra (see Methods, subsection \nameref{sec:discwinds_methods}). For the radial velocities of our variable absorption lines, roughly between $-25$ and $-115$ km/s, sodium winds are expected to have column densities ranging from $4.4 \times 10^{9}$ to $1.8 \times 10^{10}$ cm$^{-2}$ and from $1.6 \times 10^{10}$ to $6.6 \times 10^{10}$ cm$^{-2}$ assuming the minimum and maximum mass accretion rates, respectively. In contrast, the column densities measured in our sodium lines are at least nearly two orders of magnitude higher (Fig. \ref{fig6}a).\\

The above estimates assume winds to be diffuse, continuous flows of gas that effectively cover the whole stellar disc, as typically observed in disc winds \citep{Edwards+06}. However, this is at odds with our inference that the sodium absorption lines in PDS 70 often have surface fill fractions below unity (Fig. \ref{fig6}b), indicating that the gas that forms these lines is spatially confined, composed of clumps of material crossing our line of sight, instead of a large-scale, diffuse wind. Including our measured surface fill fractions in the above estimate suggests slightly higher column densities that range from $6.3 \times 10^{9}$ to $6.6 \times 10^{10}$ cm$^{-2}$ and from $2.3 \times 10^{10}$ to $2.4 \times 10^{11}$ cm$^{-2}$ assuming the minimum and maximum mass accretion rates, respectively.\\

We note that this model contains simplifying assumptions and includes parameters for which there is significant uncertainty. Given that some model parameters (including the stellar accretion rate) are significantly uncertain, this simple model prevent us from definitively ruling out the disc wind scenario for our variable absorption lines of Na\,I based solely on mass loss rate estimates. However, additional characteristics observed in our variable Na\,I absorption lines also disfavour the disc wind scenario, and we continue this discussion by comparing our variable absorption lines with previous observations of disc winds in other T Tauri stars.\\

\subsection*{Variability in other T Tauri stars}
\label{sec:variability_other_ttauri}

Absorption lines associated with disc winds are sometimes observed to vary gradually in amplitude over time, but are typically stationary in radial velocity over long periods \citep{Alencar+05}. The absence of time-variability in the radial velocity of wind absorption lines is consistent with smooth, continuous flows of material crossing our line of sight. An example of a T Tauri star that shows blue-shifted H$\alpha$ absorption attributed to  winds is the K5.5 spectral type GM Aurigae. Here, a spectral time-series spanning 22 days shows a disc wind absorption feature in the H$\alpha$ line that slowly varies in amplitude from day to day, resulting in a global change in the H$\alpha$ absorption after several days \citep{Espaillat+21}. However, this wind component is roughly stationary, presenting no signs of extreme variability in radial velocity. Another example is the K7 T Tauri star AA Tauri, that shows disc winds in He\,I, H$\alpha$, and in Na\,I \citep{AlencarBasri00, Edwards+06}, the latter tracing even cooler disc regions further away from the star. The disc wind mass loss rate of AA Tau \citep{Edwards+06, Esau+14} is measured to be 0.1--0.25 times its accretion rate of $3 \times 10^{-10}$ to $7 \times 10^{-9}$ M$_\odot$/yr \citep{Donati+10, Bouvier+13}, consistent with the expectation for disc winds of approximately 0.1--0.3 \citep{NattaGiovanardi90, Calvet97, Rigliaco+13, IguchiItoh16}. However, these sodium wind lines in AA Tau are also roughly stationary in radial velocity at approximately $-20$ km/s (Supplementary Fig. \ref{supfig1}). We note that both GM Aur and AA Tau are much younger T Tauri stars, with estimated ages around 2 Myr \citep{Espaillat+21, Bouvier+99}, and hence have higher accretion rates and higher gas densities in their protoplanetary discs. PDS 70 is significantly older, which leads to the expectation that the mass loss in the form of winds should be much less vigorous.\\

We recall that the HARPS dataset of PDS 70 presents a stationary Na\,I absorption line at $-22$ km/s across all epochs, which could be an example of a disc wind in sodium equivalent to AA Tau (Supplementary Fig. \ref{supfig1}). The same absorption line appears to be broader in the 2018 data in comparison to the 2020 data (Fig. \ref{fig2}), which could indicate either a global change in the column density of the disc wind between 2018 and 2020, or the presence of low-velocity variable Na\,I absorption lines overlapping with the line at $-22$ km/s. The modulated variation in line depth of the $-22$ km/s absorption with stellar rotation and with the emission (accretion) line may be an additional indicator that the $-22$ km/s Na\,I absorption feature originates from disc winds. However, our fitting may be affected by the proximity of the $-16$ km/s absorption line, and future analyses are necessary to properly break degeneracies between the $-22$ and the $-16$ km/s lines (see Methods, subsection \nameref{sec:stationary_lines_methods}).\\

As opposed to spatially diffuse disc winds, a rapid variability over time in the radial velocity of wind absorption lines would suggest inhomogeneities in the continuous wind outflow, caused by substructures in the circumstellar disc. Although uncommon, this is the case of RY Tauri and SU Aurigae, where a day-to-day variability in radial velocity has been detected in the H$\alpha$ and Na\,I lines \citep{Petrov+21, Petrov+23}. However, this variability is observed to be modulated in phase with a period of 255 days for SU Aur and 22 days for RY Tau, suggesting the line of sight intercepts these density regions periodically as the disc rotates. This indicates that even clumpy disc winds would produce spectral lines that periodically appear, last for a period of several days, and disappear. On the other hand, our variable absorption lines of Na\,I do not show any sign of periodicity or phase modulation, appearing and disappearing on a night-to-night timescale (Fig. \ref{fig7}). Moreover, disc wind lines in RY Tau and SU Aur are seen with the same periodicity in lines of Na\,I D and H$\alpha$. Yet, in PDS 70, disc winds are not visible in H$\alpha$, which imply that significant winds are also not expected to emanate from colder regions traced by Na\,I \citep{NattaGiovanardi90, KwanFischer11}. We conclude that only the stationary Na\,I absorption line at $-22$ km/s compares favourable to disc wind lines observed in RY Tau and SU Aur, and that the rapid appearance and disappearance of our variable Na\,I absorption lines in PDS 70 is challenging to explain by a disc wind. It is possible, however, that some of our variable Na\,I absorption lines are indeed periodic, yet a strong contamination of additional variable lines prevents us from detecting any periodicity.\\

Further comparisons can be drawn between PDS 70 and the 25 Myr old RZ Piscium system, in which variable sodium absorption has also been detected \citep{Potravnov+17}. Sodium lines in RZ Psc appear to be nearly saturated, strongly blue-shifted ($-20$ to $-120$ km/s), and variable on a day-to-day timescale, comparable to the sodium lines in PDS 70 observed in this work. The same behaviour is also seen in the Ca\,II lines, suggesting a common origin between Na\,I and Ca\,II \citep{Potravnov+17}. Classical wind models do not reproduce this stochastic variability, and therefore these authors have associated these sodium lines with a magnetic propeller effect, as a result of interactions between the circumstellar disc and the stellar magnetosphere \citep{Grinin+15}. This therefore leads to the suggestion that a magnetic propeller could be the cause of our observed sodium lines. However, the magnetic propeller theory predicts that the sodium absorption lines should be correlated with the stellar rotation period, which is not observed in our data (Fig. \ref{fig7}). In addition, the magnetic propeller occurs when the stellar accretion flow is reversed and directed into an outflow by passing magnetic field lines \citep{Grinin+15}, yet PDS 70 has a strong accretion line that is independent of the presence of additional sodium absorption lines, and an accretion rate that is much higher than RZ Psc, at the upper edge of what would be allowed for a magnetic propeller (on the order of 10$^{-10}$ M$_{\odot}$/yr). Similar to PDS 70, the short timescale, stochastic variation in the blue-shifted sodium lines in RZ Psc has already been presented as a challenge to the magnetic propeller model \citep{Potravnov+17}, while a recent study reports the presence of evaporating exocomets in its Transiting Exoplanet Survey Satellite (TESS) light curve \citep{Gibson+25}, consistent with previous studies that observe transiting, evaporating rocky material \citep{Punzi+18}. We hypothesise that the variable blue-shifted sodium lines observed in RZ Psc may be caused by exocomets, rather than the proposed magnetic propeller process.\\

\subsection*{Exocomets as the origin of the variable sodium lines}

After the above evaluation of the disc wind scenario in the spectra of PDS 70, our study explores sublimation from exocomets as the origin of the stochastically variable Na\,I absorption lines. The properties of these lines, including their rapid and stochastic radial velocity variability, the acceleration and deceleration with respect to the star, the limited fill fractions indicating confined clumps, and the relatively high column densities of this material, all support exocomet activity as a plausible interpretation. In fact, assuming the abundance of sodium in CI Chondrite meteorites, which is measured to be approximately $5.01 \times 10^{-3}$ \citep{Lodders03}, the minimum sodium masses estimated from our variable absorption lines are consistent with planetesimals with masses between $2.5 \times 10^{10}$ and $7.8 \times 10^{11}$ kg. Assuming spherical planetesimals with a rocky bulk density of 1190 kg/m$^3$, corresponding to asteroid (101955) Bennu \citep{Scheeres+19, Barnouin+19}, exocomets with diameters between roughly 0.3 to 1.1 km could already account for the minimum masses of sodium measured in the variable lines reported here, if they would be sublimated in their entirety (see Methods, subsection \nameref{sec:planetesimal_size_methods}). Furthermore, recent 2026 observations with UVES reveal the presence of low-velocity, red-shifted Na\,I absorption lines (e.g. on 2026-01-18 and 2026-03-22, see Fig. \ref{fig3}), indicating that the variable sodium lines are not exclusively blue-shifted, which provides additional evidence in favour of the exocomet scenario.\\

We compare our findings with the extensively studied 20 Myr $\beta$ Pictoris system \citep{BinksJeffries14, MamajekBell14, Shkolnik+17}. In $\beta$ Pictoris, various absorption lines with the same characteristics have been used to infer exocometary activity in metals \citep{Ferlet+87, Kiefer+14, WelshMontgomery16, Vrignaud+24, VrignaudLecavelierdesEtangs24, Vrignaud+25}, in particular on thousands of occasions in the Ca\,II H\&K lines at 396.85 and 393.36 nm \citep{Kiefer+14, Hoeijmakers+25}. In $\beta$ Pictoris, absorption lines of ionised species are more commonly observed than absorption lines of the same neutral species. This can be explained by the fact that $\beta$ Pictoris is a hot A-type star, and the gas released from exocomets grazing the star is exposed to the strong stellar UV field, which efficiently ionises metal species. As a consequence, in the case of calcium, its ionised state becomes more abundant than its neutral state, which facilitates the detection of Ca\,II in comparison with Ca\,I. However, the PDS 70 host star is much cooler than $\beta$ Pictoris, and we infer that the material sublimating from exocomets in PDS 70 are less strongly ionised due to the lower effective temperature of the star and a redder spectral energy distribution. Even if the exocomet clouds may be less subject to ionisation, exocomets in PDS 70 could still be able to generate significant absorption by the near-UV Ca\,II H\&K lines, and we investigated the HARPS spectra at these wavelengths. However, the flux at these wavelengths is so low around the Ca\,II H\&K doublet that the stellar continuum is essentially not detected. In addition, the Ca\,II H\&K doublet in PDS 70 has a very strong, possibly chromospheric emission line \citep{Thanathibodee+20}, which hinders the detection of any variable stellar absorption lines. Neutral species such as Na\,I are less abundant than its ionised state around $\beta$ Pictoris, explaining why Na\,I is only rarely observed in $\beta$ Pictoris exocomets \citep{Hoeijmakers+25}. On the other hand, Na\,I that sublimates from exocomets is expected to survive more easily around PDS 70, motivating the exocomet phenomenon to be attributed to sodium absorption.\\

\subsection*{Blue-shifted exocomet lines}
\label{sec:blueshifted_exocomet_lines}

The variable sodium lines observed in the 2018 spectra of PDS 70 are preferentially blue-shifted. Under the assumption that these clouds trace the orbital motion of the exocomet around the star, non-zero Doppler-shifts in transit require highly eccentric orbits. A blue-shift implies that the object is approaching the observer along the line of sight, meaning that the distance between it and the star must be increasing. The high radial velocities observed for components suggest that these exocomet clouds are likely near periastron. This means that these objects have passed their closest approach to the star, and are thus now receding from periastron. Near periastron, the sublimation of material from the parent body is expected to be at its maximum, and also the transit probability is highest for orbits where the periastron is near our line of sight \citep{Kane+12}.\\

The 2026 dataset shows that, even in the presence of red-shifted Na\,I absorption lines (Fig. \ref{fig3}), most variability is still seen in the blue-shifted side of the spectra. In the exocomet scenario, a preference for the occurrence of blue-shifted components implies that the exocomets are part of a population of objects with similar orbital characteristics. This is known to be the case in the $\beta$ Pictoris system, where exocomets are preferentially red-shifted because the population is in mean-motion resonance with the planet $\beta$ Pictoris c \citep{Lacour+21, Beust+24, JaworskaHoeijmakers26}. In $\beta$ Pictoris, this preference is expected to be transient, as the precession of the orbit of $\beta$ Pictoris c results in a gradual modulation of the arguments of periastron of the entire exocomet population \citep{Beust+24}. In PDS 70, however, we presently have no knowledge of additional planets in the inner system that might structurally affect the dynamical evolution of planetesimals there.\\

On some occasions (e.g. on 2018-03-29 and 2018-03-31), changes in the best-fit radial velocity on the order of 1 km/s are apparent. Such acceleration can be due to the curvature of the trajectory in parts of the orbit very close to periastron, and this effect has also been observed in $\beta$ Pictoris \citep{Kennedy18}. However, on some occasions, the acceleration results in an increasing blue-shift, and this is impossible for Keplerian motion. Similar behaviour has also been observed in $\beta$ Pictoris and has been attributed to time-varying dynamics in the evaporating tail \citep{Hoeijmakers+25}. It is possible that there is a significant discrepancy between the radial velocities of the sublimated tails streaming towards the observer, and the orbital motion of the sublimating exocomets themselves. It is therefore unclear if significant alignment of the orbital parameters of the exocomet population akin to the mean-motion resonance phenomenon in $\beta$ Pictoris is required to explain these observations, and more simulations of tail dynamics in the environment of PDS 70 would be required to estimate the magnitude of the tail velocity compared to the orbital velocity.\\

\subsection*{Exocomet scattering from the outer disc towards the star}
\label{sec:exocomet_scattering}

In PDS 70, there are two known gas giant planets that are observed to have cleared a gap in the protoplanetary disc \citep{Hashimoto+12, Hashimoto+15, Keppler+18, Muller+18, Benisty+21}. These planets could be a potential source of planetesimals on stargrazing orbits, although their masses are still poorly constrained, with uncertainties of several Jupiter masses \citep{Haffert+19, Mesa+19}. Alternatively, the existence of exocomets may be evidence for the presence of additional planets in the inner system. Indeed, high-contrast imaging observations with the Spectro-Polarimetric High-contrast Exoplanet REsearch (SPHERE) at the Very Large Telescope (VLT) \cite{Mesa+19, Hammond+25} as well as recent JWST observations \citep{Christiaens+24} suggest the possible presence of a third gas giant planet interior to the orbit of PDS 70 b in the inner disc, at a distance of around 13.5 au from the star. Yet, we note that the tally of planets interior to the newly proposed PDS 70 d may still be incomplete, and we hypothesise that an unseen planet closer to the star could potentially be responsible for scattering exocomets, similar to the interactions observed in $\beta$ Pictoris c. We recall that the existence of the planets in the $\beta$ Pictoris system was proposed based on such dynamical simulations of the exocomet phenomenon, over 10 years before $\beta$ Pictoris b was ultimately discovered \citep{Beust+90, Levison+94, Beust96}. Additional possibilities are that the exocomets in PDS 70 could originate from a single parent body that broke up due to tidal disruption after close encounter with a planet \citep{Opik66, Marsden67, Beust96}.\\

An evidence that favours the delivery scenario is that PDS 70 is a low accreting T Tauri star, yet its mass accretion rate is higher than expected from the age of the system, which requires that the mass reservoir of the inner disc of PDS 70 is resupplied \citep{Thanathibodee+20}. Observations from the Atacama Large Millimeter/submillimeter Array (ALMA) indicate that gas and dust are depleted in the gap between the inner and outer discs \citep{Keppler+19, Facchini+21}, which may effectively hinder the influx of material. However, dust evolution models of the system have shown that particles smaller than 0.1 $\upmu$m are able to be transported from the outer disc, surviving accretion by PDS 70 c and b, and feeding the inner disc \citep{Pinilla+24}. An unexplored hypothesis in the delivery scenario is if planetesimals from the outer disc could also be able to resupply the mass of the inner disc, delivering water and volatiles to the stellar vicinity, as proposed in this work. This scenario is particularly interesting considering the detection of water vapour in the inner disc, at a radial distance of 0.05 au from the star \citep{Perotti+23}, which could require a delivery mechanism.\\

We have carried out N-body dynamical simulations using \texttt{REBOUND} \citep{ReinHu12} to investigate whether planetesimals from the outer disc are prone to being scattered inwards via interactions with the planets. We performed two sets of simulations including two planets (PDS 70 b and c) or three planets (PDS 70 b, c, and candidate d). Our simulations assume planetary parameters based on previous N-body simulations of the two- and three-planet cases, assuming stability of the planetary orbits \citep{Trevascus+25} (see Methods, subsection \nameref{sec:nbody_simulations_methods}). Since the presence of PDS 70 d affects the overall stability of the system, the two- and three-planet cases use different parameters, in particular the mass of PDS 70 c, which is higher in the two-planet case.\\

We expected that the outermost planet PDS 70 c would have the most influence in the scattering of particles from the outer disc, as it is the most massive and the closest to the outer disc. The difference between the assumed mass of PDS 70 c in the two- and three-planet cases results in a large fraction of particles being scattered in the first 1 Myr of the simulation in the two-planet case (Fig. \ref{fig8}b), as opposed to a more smooth scattering of particles over 5 Myrs in the three-planet case (Fig. \ref{fig8}e). Nevertheless, our simulations show that particles scatter inwards in both cases, which means that PDS 70 c does so efficiently even in the low-mass scenario. In the last 1 Myr, an average of 0.47 and 2.9\% of particles cross the gap towards the inner disc in the two- and three-planet case, respectively. Our simulations therefore indicate that, for every 10 M$_{\oplus}$ of planetesimals in the outer disc at the current time, gravitational interactions with the planets may scatter 0.047 and 0.29 M$_{\oplus}$ across the gap per Myr in the first and second cases, respectively. The total planetesimal masses contained in the inner and outer discs are currently unknown, but based on ALMA observations, the dust mass is estimated to be 0.08--0.36 M$_{\oplus}$ in the inner disc \citep{Benisty+21, Fasano+25} and at least 28 M$_{\oplus}$ in the outer disc \citep{Doi+24}. Our simulations therefore support the notion that a significant number of planetesimals may efficiently evolve from the outer disc to the regions interior to the gas giants, with or without the presence of the newly proposed planet PDS 70 d. These planetesimals may subsequently become available to feed the inner region with volatiles, dust \citep{Pinilla+24}, and stargrazing exocomets.\\

\subsection*{Orbital eccentricities and inclinations of transiting exocomets}
\label{sec:exocomet_inclinations_eccentricities}

It is expected that planetesimals in young systems are born in orbits close to the mid-plane, with low inclinations with respect to the disc. For this reason, exocomets in transit would intuitively be more easily detectable in highly inclined, edge-on system such as $\beta$ Pictoris \citep{Augereau+01, Milli+14}. The disc of PDS 70 is misaligned, with an inclination of $51.7 \pm 0.1\degree$ relative to the plane of the sky \citep{Keppler+19}, which could disfavour the observation of transiting exocomets. However, N-body simulations of $\beta$ Pictoris have shown that the orbits of long-period exocomets tend to acquire a broad range of inclinations with respect to the disc, after repeated interactions with the two planets in that system \citep{JaworskaHoeijmakers26}, which suggests that an edge-on orientation is not a requirement for observing the transit of exocomets that originated in the outer disc.\\

In PDS 70, assuming that planetesimals initially orbit the star in the plane of the disc, planetesimals should have obtained orbital inclinations of $38.3\degree$ or higher to be able to transit. Our N-body simulations demonstrate that most planetesimals that cross the gap towards the inner disc retain inclinations close to the plane of the disc. Yet, about 10\% of the planetesimals that cross the gap dynamically evolve onto orbital inclinations higher than $38.3\degree$ up to $180\degree$ relative to the plane of the disc after interactions with PDS 70 b and c (Fig. \ref{fig8}b), particularly the ones that evolve onto orbits that are entirely within the orbit of PDS 70 c. In our N-body simulation that includes the presence of PDS 70 d, the fraction of planetesimals that evolve onto orbital inclinations higher than $38.3\degree$ is 6\% (Fig. \ref{fig8}f), which is lower than the two-planet simulation due to the lower assumed planet masses, yet still significant. Importantly, these inclined orbits are also highly eccentric, in accordance with the exocomet hypothesis (Fig. \ref{fig8}c and Fig. \ref{fig8}g). Our simulations therefore suggest that, in principle, the orbits of a significant number of planetesimals can indeed achieve sufficiently inclined and eccentric orbits as a result of dynamical interactions with the planets, particularly with PDS 70 c.\\

We emphasise, however, that these estimates are insufficient to provide a direct comparison with the frequency of exocomets inferred from the HARPS observations, due to the high uncertainties of the constituents of the system, in terms of planet masses, orbital properties as well as the number of planets. In particular, the existence of any additional inner planets could dramatically change the nature of the dynamical evolution of particles in the inner system, exciting the inclination distribution of scattered planetesimals even further. Instead, these simulations primarily show that transport of material from the outer regions of the system and across the orbits of planets PDS 70 c, b, and even d is plausible, establishing migration of planetesimals as an alternative channel of resupplying the inner disc of PDS 70.\\

\bigskip
\bigskip
\bigskip

In this study we investigated variable Na\,I absorption lines in the high-resolution HARPS spectra of PDS 70. Even being preferentially blue-shifted, we find that these are challenging to explain using models of disc winds, and instead we propose that the properties of these lines form plausible evidence for transiting exocomets in the PDS 70 system. Red-shifted variable Na\,I absorption lines present in UVES observations from 2026 also support the exocomet scenario in PDS 70. Additionally, we carried out N-body simulations of the gravitational interactions of planetesimals with the planets in the system, and find that these are able to scatter planetesimals from the outer disc beyond 54 au towards the inner system within the first few million years of the system, supporting the hypothesis of planetesimal transport as a viable mechanism to supply the inner disc of volatiles \citep{Perotti+23}, as well as placing planetesimals on highly eccentric, stargrazing orbits. If confirmed, PDS 70 is the youngest system in which the exocomet phenomenon has been observed \citep{Ferlet+87, MontgomeryWelsh12, Kiefer+14, Kiefer+14b, Rappaport+18, Strom+20, Iglesias+25}, and only the second system with exocomet activity where planets have also been confirmed \citep{Keppler+18, Muller+18} --- the first being the reference system $\beta$ Pictoris. PDS 70 would also be the coolest star with exocomets, which would imply that the exocomet phenomenon is not exclusive to early-type stars \citep{Strom+20, Dumond+25, Gibson+25}.\\

In order to unambiguously confirm exocomet activity in PDS 70, we believe that subsequent observations would be needed, including: i) photometric observations of transits of the dust tails of the PDS 70 exocomets, which is hampered by the stellar variability; ii) time-series observations to constrain the minimum variability timescale (ingress, egress, and typical transit duration), to resolve acceleration (variation of radial velocity with time) of individual exocomets to constrain orbital parameters of individual comets, in particular their periastron distance \citep{Kennedy18}; iii) measurements of elemental abundance ratios of species with varying volatility \citep{Vrignaud+25}, to test that the material has not been primordially devolitalised close to the star, iv) better characterisation of the system, including planetary masses and detection of unseen planets, which is expected to become significantly more tractable with the Extremely Large Telescope (ELT). New, direct observations of exocomets in PDS 70 will provide a unique observational window into the properties of solid, planet-building material at an age where accretion on both the planets and the protostar are still ongoing \citep{Thanathibodee+20, Benisty+21}. Furthermore, the presence of sublimating exocomets combined with the recent JWST/MIRI observations \citep{Perotti+23} could be evidence for an ongoing episode of volatile delivery by planetesimals from the outer disc to the inner, terrestrial-planet region of the PDS 70 system. Sustained study of exocomets in PDS 70 may therefore bear relevance for theories of water delivery to the early inner Solar System \citep{OwenBarNun95, Obrien+06, Raymond+09, Walsh+11, Obrien+14, RaymondIzidoro17, OBrien+18, Mandt+24}.\\

\section*{Methods}
\label{sec:methods}

\subsection*{Observations and data analysis}
\label{sec:observations_analysis_methods}

HARPS is a fiber-fed, cross-dispersed echelle spectrograph with a resolving power of $R = 115\,000$ covering the optical wavelength range from 380--690 nm. The archival data analysed in this work include 52 spectra, spanning three years of sporadic monitoring of the system in the first semesters of 2018, 2019, and 2020 (Programme ID 098.C-0739, 101.C-0557, 1101.C-0557, 0104.C-0418, PI A.-M. Lagrange). A detailed observation log can be found in Table \ref{table1}. Observations were taken with fiber A on the target and fiber B on thorium-argon (ThAr) lamps (2018-03-29, 2018-03-30, and 2018-03-31) or on an empty field on the sky (remaining spectra). We downloaded the processed (wavelength calibrated and reduced) science spectra from the ESO Archive database.\\

We used the \texttt{molecfit} software \citep{Smette+15, Kausch+15} to generate telluric models of \ch{H2O} and \ch{O2} for the 52 HARPS spectra. We used the Cross-correlation analysis of high resolution spectroscopy tool \texttt{Tayph} \citep{Hoeijmakers+20, Prinoth+22, Borsato+23} to read the 2D extracted spectra (e2ds) in all 72 orders. \texttt{Tayph} converts each spectrum from air to vacuum wavelengths, and corrects for the barycentric Earth radial velocity before feeding it into \texttt{molecfit}. After generating telluric models for each file, for all orders, we divided the flux by the corresponding telluric model (Fig. \ref{fig2}). The same telluric correction procedure was done for the UVES data obtained in 2026 (Programme ID 0116.C-0329, PI A. Novais, Fig. \ref{fig3}).\\

For spectra taken with fiber B on the sky, we corrected emission lines from sodium in the Earth's atmosphere by subtracting the sky flux from fiber B from the corresponding spectral flux obtained from fiber A. For spectra taken with fiber B illuminated with the ThAr lamp, we removed the data points approximately between 589.167--589.183 nm and 589.767--589.782 nm, where Na\,I emission sky lines are expected. We then shifted each spectrum to the stellar rest frame by correcting for the measured system radial velocity of 6.01 km/s, derived using the same HARPS data (see Methods, subsection \nameref{sec:vsys_methods}). For the UVES data, we also removed the data points approximately at the location of the Na\,I emission sky line and shifted the spectra to the stellar rest frame.\\

\subsection*{Fitting sodium doublet lines}
\label{sec:fitting_methods}

Of the 12 nights of observation in 2018, most of these had at least two spectra taken per night (see Table \ref{table1}). Each of the first three nights (2018-03-29, 2018-03-30, and 2018-03-31) had 6 spectra available, which revealed signs of absorption line variability within the duration of the night (Fig. \ref{fig2}). In order to increase the signal-to-noise ratio for each of these nights for the fitting procedure, yet making sure this variability was still captured, we paired the 6 exposures observed on the first three nights in three pairs per night. The remaining 8 nights had a maximum of two observations per night, and no significant variability is seen between spectra of the same night. Therefore, spectra from nights with only one observation were fitted singly, while spectra from nights with two observations were paired, resulting in a total of 18 epochs. We fitted the photospheric Na\,I D lines for each of the 18 epochs separately, consistent with the methodology of Kiefer et al. \citep{Kiefer+14}.\\

We follow the line fitting procedure detailed in Hoeijmakers et al. \citep{Hoeijmakers+25}, which follows the analytical model of Kiefer et al. \citep{Kiefer+14} assuming a homogeneous cloud of absorbing sodium gas. To fit the Na\,I doublet lines, we slice the combined spectra in two separate wavelength sections that are $\pm$135 km/s from each of the two Na\,I D lines (588.893--589.424 nm for D$_2$ and 589.490--590.021 nm for D$_1$). These ranges allow us to include all variable Na\,I absorption components in our fitting region while also avoiding absorption lines of other species. We then normalise both spectral sections simultaneously to the median flux of both sections to preserve relative flux ratios. We then fit each photospheric Na\,I D line as a broad absorption component with a pseudo-Voigt profile at the stellar rest frame. Superimposed in each broad photospheric absorption line, we simultaneously fit: two Gaussian absorption components at roughly $-22$ and $-16$ km/s; a Gaussian emission component at the stellar rest frame; and 1--4 additional Gaussian absorption components that vary across epochs. Each component on the first spectral section (around Na\,I D$_2$) will have their doublet counterpart on the second section (around Na\,I D$_1$) with an optical depth ratio of approximately 2, matching the ratio of oscillator strengths. All line components in both sections are fitted simultaneously.\\

For low optical depths ($\tau < 1$, where $\tau$ is the optical depth measured at line centre), the first sodium absorption line (D$_2$) is expected to be about twice as strong as the second line (D$_1$), meaning their observed line ratio is approximately equal to 2. However, if both doublet lines are saturated (i.e. $\tau > 1$), their line ratio may approach 1. By fitting both doublet lines simultaneously, we are able to determine the optical depth of the lines in the optically thick limit \citep{Hoeijmakers+25}. Furthermore, we assume that the stellar disc may not be fully covered by the cloud by accounting for a surface fill fraction $f$, i.e. the fraction of the stellar disc covered by the cloud of gas: If part of the stellar light is not absorbed by the gas, the observed flux is diluted by the unabsorbed stellar light. Fitting the surface fill fraction $f$ allows us to determine the true optical depth, despite this dilution \citep{Hoeijmakers+25}. We note, however, that the surface fill fraction does not necessarily provide a measurement of the gas cloud radius, unless one assumes that the cloud is entirely projected onto the stellar disc.\\

Each absorption line component (with the exception of the photospheric absorption line) is fitted with 4 free parameters: the centre wavelength of the first doublet line $\mu$ (in nm); the Gaussian line width of the first doublet line $\sigma_v$ (in km/s); the optical depth at the centre of the first line $\tau$; and the surface fill fraction of the projected stellar disc $f$. The photospheric absorption line and the emission line have $\mu$, $\sigma_v$, and $\tau$ as free parameters, but their surface fill fractions $f$ are fixed to 1. The photospheric absorption line is poorly modelled by a Gaussian shape, and therefore also has a Lorentzian line width parameter $\gamma_v$ (in km/s). An additional free parameter in our model is the separation between the first and the second doublet lines $d\mu$ (given in nm), which is assumed to be the same for all components in each spectrum. We assume a linear function to fit the spectral continuum at each section with the polynomial coefficients $c_0$ and $c_1$ as free parameters for the first section (around Na\,I D$_2$) and $d_0$ and $d_1$ for the second section (around Na\,I D$_1$). Each epoch will therefore have between 24 and 36 free parameters, depending on the number of variable lines.\\

Therefore, given an absorption line component, the flux of the first and second doublet lines are given by

\begin{equation}
    F_{{\rm A}_{\rm D_2}}(v) = 1 - f \left[ 1 - e^{-\tau \Phi(v)} \right] \text{\,\,\, and \,\,\,} F_{{\rm A}_{\rm D_1}}(v) = 1 - f \left[ 1 - e^{-\frac{\tau}{R} \Phi(v)} \right] \text{,}
\end{equation}\\

\noindent
respectively, where $R$ is the ratio between the oscillator strengths of the doublet lines, which for Na\,I D is approximately equal to 2 \citep{Forbrich+68} and is kept fixed, and $\Phi(v)$ is the line profile. For all lines except for the photospheric absorption line, we use a Gaussian profile $\Phi_{\rm G}(v)$, which is given by

\begin{equation}
    \Phi_{\rm G}(v) = e^{-\frac{1}{2} \left( \frac{v}{\sigma_v} \right)^2} \text{\,\,\, where \,\,\,} v(\lambda) = \frac{c}{\mu}(\lambda - \mu)
\end{equation}\\

\noindent
and $\sigma_v$ is the Gaussian standard deviation. For the photospheric absorption line, we use a pseudo-Voigt profile $\Phi_{\rm V}(v)$, as described by \citep{Thompson+87}, where $\gamma_v$ is the Lorentzian half-width at half-maximum.\\

The flux of the first and second doublet emission lines are given by

\begin{equation}
    F_{\rm E_{\rm D_2}}(v) = \tau \Phi(v) \text{\,\,\, and \,\,\,} F_{\rm E_{\rm D_1}}(v) = \frac{\tau}{R} \Phi(v) \text{,}
\end{equation}\\

\noindent
respectively, and the flux of the continuum is given by

\begin{equation}
    F_{\rm C_{\rm D_2}}(v) = c_0 + c_1 v \text{\,\,\, and \,\,\,} F_{\rm C_{\rm D_1}}(v) = d_0 + d_1 v
\end{equation}\\

\noindent
for both spectral sections. The flux of each spectral section is modelled as

\begin{equation}
\begin{gathered}
    F_{\rm D_2}(\lambda) = \left[ F_{\rm A_{{\rm ph}_{\rm D_2}}}(v) F_{{\rm C}_{\rm D_2}}(v) + F_{{\rm E}_{\rm D_2}}(v) \right] F_{{\rm A_1}_{\rm D_2}}(v) ... F_{{{\rm A}_N}_{\rm D_2}}(v) \\
    \text{\,\,\, and \,\,\,} \\
    F_{\rm D_1}(\lambda) = \left[ F_{\rm A_{{\rm ph}_{\rm D_1}}}(v) F_{{\rm C}_{\rm D_1}}(v) + F_{{\rm E}_{\rm D_1}}(v) \right] F_{{\rm A_1}_{\rm D_1}}(v) ... F_{{{\rm A}_N}_{\rm D_1}}(v)
\end{gathered}
\end{equation}\\

\noindent
where $F_{\rm A_{\rm ph}}(v)$ is the photospheric absorption line. The total flux is then determined as the mean value between the first and second doublet fluxes for all line components:\\

\begin{equation}
    F(\lambda) = \frac{F_{\rm D_2}(\lambda) + F_{\rm D_1}(\lambda)}{2} \text{.}
\end{equation}\\

We use \texttt{NumPyro} \citep{Phan+19, Bingham+19} to implement a No U-Turn Sampler (NUTS) \citep{Betancourt17, HoffmanGelman14}, which is a Markov chain Monte Carlo (MCMC) algorithm that optimises the sampling by automatically adapting the number of steps per iteration, avoiding retracing explored steps while making sure the parameter space is effectively explored. The prior ranges used for each line component in each of the 18 epochs are listed in Supplementary Tables \ref{suptable1} and \ref{suptable2}. The best-fit results for each line component of each epoch are listed in Supplementary Tables \ref{suptable3} and \ref{suptable4}. Supplementary Tables \ref{suptable1}--\ref{suptable4} are also available electronically (see \nameref{sec:data_availability}).\\

\subsection*{Derivation of the systemic velocity}
\label{sec:vsys_methods}

To determine the systemic velocity of the PDS 70 system, we cross-correlated all spectra after barycentric correction up to spectral order 30 with a PHOENIX model spectrum ($T_{\rm eff} = 4152$ K, $\log g = 3.68$ cm\,s$^{-2}$ \citep{Swastik2021}). Higher orders were excluded to avoid telluric contamination. Using \texttt{PyMultiNest} \citep{Buchner+14}, we then fitted a rotationally broadened Voigt profile \citep{Gray2008} to the average cross-correlation function, assuming a limb-darkening coefficient of $E = 0.6$, appropriate for optical wavelengths \citep{Thanathibodee+20, Claret2000}. The fit was restricted to the 2018 observations, as only these were used in the analysis. We find a systemic velocity of $6.01^{+1.91}_{-2.18}$ km/s (Supplementary Fig. \ref{supfig2}), consistent with the value reported by Thanathibodee et al. \cite{Thanathibodee+20} using the same HARPS dataset. All spectra were corrected to the stellar rest frame using this velocity, without accounting for its uncertainty. However, we note that the variable Na\,I absorption lines are significantly more blue-shifted than the uncertainty in the systemic velocity (Fig. \ref{fig1}b).\\

\subsection*{Stationary sodium lines}
\label{sec:stationary_lines_methods}

The outcomes of our fitting procedure show that the line centre positions of the photospheric absorption and emission lines are both consistently stationary are at roughly the stellar rest frame (Fig. \ref{fig7}). The peak flux of the emission line appears to vary periodically with the stellar rotation period of $3.03 \pm 0.06$ days, which has been confidently measured using the TESS light curve obtained in April 23--May 20, 2019 as part of Sector 11 \citep{Thanathibodee+20}. A similar modulation with the stellar rotation period has been observed in H$\alpha$ in previous studies \citep{Thanathibodee+20}, in the same HARPS dataset acquired in 2018, fitted using magnetospheric accretion models. The H$\alpha$ accretion lines are variable in amplitude, which we interpret as corresponding to increases or decreases in the mass accretion rate. The H$\alpha$ line also exhibits a chromospheric emission component, with radial velocity at the line centre near the velocity of the stellar rest frame. Based on these findings for H$\alpha$, we conclude that the nature of the Na\,I emission line in the spectra of PDS 70 is likely due to accretion of material onto the star.\\

The two narrow, partially blended Na\,I absorption components were measured to be stationary at approximately $-22$ and $-16$ km/s across all epochs (Fig. \ref{fig7}). The depths of both absorption lines vary throughout the observations; however, their variation does not appear to be correlated with each other, which could indicate that they have different origins. Here, we note that the blending between the two components in some epochs may significantly affect their fitting. When the components are partially overlapping, their parameters become degenerate, which may overestimate the width or optical depth of one of the lines, suppressing the other. The emission line together with the $-22$ km/s absorption line resembles a P Cygni profile, as commonly seen in disc winds lines of Na\,I \citep{NattaGiovanardi90}, He\,I \citep{CampbellWhite+23}, and other species \citep{AlencarBasri00}. Due to this modulation with stellar phase and time, we attribute the $-22$ km/s stationary Na\,I absorption line as gas outflowing from the circumstellar disc (see also \nameref{sec:variability_other_ttauri}). The depth of the red-most component at $-16$ km/s appears to vary in time, although no periodicity with stellar rotation is observed. While it is possible that this component is circumstellar, originating further out in the disc, its line depth and velocity may also be compatible with absorption emerging from the interstellar medium (ISM) \citep{Welsh+10}. The line depths of ISM absorption lines are generally stable over a timescale of years \citep{McEvoy+15}, yet the blending of this component with the ISM line may significantly affect our fitting result, affecting the apparent line parameters. ISM Na\,I lines are seen in other stars of the Scorpio-Centaurus association, where PDS 70 is located. In particular $\mu$ Centauri and $\phi$ Centauri, which are at nearby line of sights (angular separations of 3.6$\degree$ and 2.0$\degree$ from PDS 70, respectively), both present ISM Na\,I lines at approximately $-10$ km/s in the heliocentric frame \citep{Crawford91}. This value is consistent with the velocity of our stationary line at $-16$ km/s (in the stellar rest frame, which converts to a heliocentric velocity of $-10$ km/s, assuming the measured systemic velocity of 6.01 km/s derived from the data, see Methods, subsection \nameref{sec:vsys_methods}), which supports the possibility that our $-16$ km/s stationary Na\,I line is likely blended with an ISM line. However, follow-up measurements are necessary to confirm the physical mechanisms that cause both the $-22$ and $-16$ km/s Na\,I absorption lines and to ascertain the importance of ISM absorption.\\

\subsection*{Calculation of sodium column density}
\label{sec:column_density_methods}

Assuming that self-shielding is negligible, the sodium column density is given by

\begin{equation}
    N_{\rm col} = \frac{\tau(\mu)}{\sigma_{\rm cs}(\mu)} \text{,}
\end{equation}\\

\noindent
where $\tau(\mu)$ is the optical depth and $\sigma_{\rm cs}(\mu)$ is the absorption cross-section of sodium, both at the centre wavelength $\mu$ of each sodium line. From \citep{RybickiLightman86}, the absorption cross-section for an atomic transition with oscillator strength $f_{\rm osc}$ is given by

\begin{equation}
    \sigma_{\rm cs}(\mu) = \frac{\pi e^2}{m_{\rm e} c^2} f_{\rm osc} \phi(\mu) \text{,}
\end{equation}\\

\noindent
where $m_{\rm e}$ is the electron mass, $e$ is the electron charge, $c$ is the speed of light, and $\phi(\mu)$ is the normalised Gaussian amplitude at line centre, given by

\begin{equation}
    \phi(\mu) = \frac{1}{\sqrt{2 \pi} \sigma_\lambda} \text{,}
\end{equation}\\

\noindent
where $\sigma_\lambda$ is the Gaussian line width. We assume the dominant mechanism responsible for broadening the profile of our exocomet sodium lines is velocity dispersion, with a Gaussian line width given by $\sigma_{\lambda} = \sigma_v \lambda_0 / c$, where $\lambda_0$ is the rest-frame Na\,I D$_2$ line wavelength in vacuum.\\

Finally, the column density of our exocomet sodium lines is given by

\begin{equation}
    N_{\rm col} = \frac{m_{\rm e} c^2}{\pi e^2} \frac{\sqrt{2 \pi} \sigma_{\lambda}} {f_{\rm osc} \lambda_0^2} \tau(\mu) \text{.}
\end{equation}\\

We note that this calculation could be used to estimate a column density for each individual line of the sodium doublet. Given that the ratio between the oscillator strengths of the first (D$_2$) and second (D$_1$) sodium doublet lines is approximately equal to 2 \citep{Forbrich+68}, in our fitting method we fix the ratio of optical depths of both lines so that the column density of sodium is fitted using both lines simultaneously. Using the radial velocity $v$, Gaussian line width $\sigma_v$, and optical depth $\tau$ measured at the centre of our 43 variable sodium absorption doublet line components, we estimate the column density of sodium to vary between $2.4 \times 10^{11}$ and $7.1 \times 10^{12}$ cm$^{-2}$ for each pair of sodium lines (Fig. \ref{fig6}a).\\

\subsection*{Calculation of sodium cloud mass}
\label{sec:cloud_mass_methods}

Assuming a spherical star with a projected circular area $A$, and assuming a homogeneous cloud of sodium gas entirely projected on the stellar disc, the mass of the sodium cloud $M_{\rm Na\,cloud}$ from each sodium line doublet is given by 

\begin{equation}
    M_{\rm Na\,cloud} = f A N_{\rm col} m_{\rm Na} \text{,}
\end{equation}\\

\noindent
where $f$ is the surface fill fraction, $N_{\rm col}$ is the sodium column density, and $m_{\rm Na}$ is the mass of the sodium atom.\\

Note that this slab model assumes that the cloud is entirely confined in front of the star, yet this may not be the case. For clouds that extend beyond the radius of the star, the inferred masses can be interpreted as lower limits. Our column densities and surface fill fractions suggest that these transiting clouds have minimum masses between $1.3 \times 10^{8}$ and $3.9 \times 10^{9}$ kg of sodium (Fig. \ref{fig6}b).\\

\subsection*{Estimation of mass loss due to disc winds}
\label{sec:discwinds_methods}

Following the methodology of \citep{Calvet97} and \citep{Thanathibodee+20}, the total mass loss rate of disc winds, which is dominated by hydrogen, can be modelled as

\begin{equation}
    \dot{M}_{\rm wind} = A v \mu m_{\rm H} n_{\rm H} \text{,}
\end{equation}\\

\noindent
where $A$ is the effective area of the wind, $v$ is the velocity of the wind, $\mu$ is the mean molecular weight, $m_{\rm H}$ is the mass of hydrogen, and $n_{\rm H}$ is the number density of hydrogen. The number density of hydrogen can be written relative to the number density and abundance of neutral sodium $n_{\rm Na}$, i.e.

\begin{equation}
    n_{\rm H} = \eta n_{\rm Na} = \eta \frac{N_{\rm col}}{L} \text{,}
\end{equation}\\

\noindent
where $N_{\rm col}$ is the column density of sodium, $L$ is the distance of the line of sight through the wind, and

\begin{equation}
    \eta = \frac{1}{X_{\rm Na} f_{\rm Na\,I}} \text{,}
\end{equation}\\

\noindent
where $X_{\rm Na}$ is the abundance of sodium relative to hydrogen and $f_{\rm Na\,I}$ is the fraction of Na\,I that is neutral, and in the ground state. Using version C25.00 of \texttt{Cloudy} (last described by \citep{Gunasekera+25}) assuming a blackbody temperature of 3972 K \citep{PecautMamajek16, Keppler+18}, luminosity of 0.35 L$_\odot$ \citep{PecautMamajek16, Keppler+18}, wind thickness $L$ of $1 R_{\rm star}$ (equal to \citep{Thanathibodee+20}, and reasonable because we find that the extent of the transiting sodium clouds tend to be equal to or smaller than the star through our fitting of the fill fraction $f$), and a distance of $10 R_{\rm star}$ to the star, where $R_{\rm star} =$ 1.26 $\rm{R}_\odot$ \citep{Keppler+18}, we adopt a sodium abundance of $X_{\rm Na} = 2.14 \times 10^{-6}$ and a fraction of neutral sodium of $f_{\rm Na\,I} = 2.65 \times 10^{-4}$, meaning that the vast majority of sodium is expected to be in an ionised state. Following \citep{Thanathibodee+20} and \citep{Calvet97}, we assume $A \sim \pi (2R_{\rm star})^2$, $\mu = 2.4$, and $\dot{M}_{\rm wind} / \dot{M}_{\rm acc} = 0.1$. Assuming our measured values for the radial velocity $v$ of each blue-shifted sodium line (Fig. \ref{fig6}a), and solving for $N_{\rm col}$, we estimate that sodium winds in PDS 70 are expected to have column densities of $4.4 \times 10^{9}$ to $6.6 \times 10^{10}$ cm$^{-2}$.\\

\subsection*{Estimation of planetesimal size}
\label{sec:planetesimal_size_methods}

The minimum sodium cloud masses of $1.3 \times 10^{8}$ and $3.9 \times 10^{9}$ kg, estimated from our variable absorption lines, can be used to estimate the size of the planetesimals from where this sodium would be sublimated. Assuming $X_{\rm Na}$ as the abundance of sodium in CI Chondrite meteorites, which is measured to be approximately $5.01 \times 10^{-3}$ \citep{Lodders03}, if the entire sodium cloud mass $M_{\rm Na\,cloud}$ was sublimated from a planetesimal, the mass of this planetesimal $M_{\rm planetesimal}$ is given by

\begin{equation}
    M_{\rm planetesimal} = \frac{M_{\rm Na\,cloud}}{X_{\rm Na}} \text{.}
\end{equation}\\

For our estimated minimum sodium cloud masses of $1.3 \times 10^{8}$ and $3.9 \times 10^{9}$ kg, these planetesimals would therefore have masses between $2.5 \times 10^{10}$ and $7.8 \times 10^{11}$ kg. Assuming a spherical planetesimal with a bulk density $\rho$, its radius corresponds to\\

\begin{equation}
    R_{\rm planetesimal} = \left( \frac{3 M_{\rm planetesimal}}{4 \pi \rho} \right)^{1/3} \text{.}
\end{equation}\\

For a rocky bulk density of 1190 kg/m$^3$, corresponding to asteroid (101955) Bennu \citep{Scheeres+19, Barnouin+19}, these minimum sodium cloud masses are consistent with planetesimals with diameters between 0.3 to 1.1 km, assuming they would be sublimated in their entirety. Analogously, for a comet-like bulk density of 533 kg/m$^3$, as of comet 67P/Churyumov-Gerasimenko \citep{Patzold+16}, these minimum sodium cloud masses could be sublimated from planetesimals with diameters between 0.5 and 1.4 km.\\

\subsection*{N-body simulations}
\label{sec:nbody_simulations_methods}

We perform N-body simulations of the PDS 70 system including the star and its planets to investigate if planetesimals (particles) originated in the outer disc are able to be scattered towards the inner system after dynamical interactions with the planets. Our simulations are performed using the non-symplectic N-body integrator \texttt{IAS15} \citep{ReinSpiegel15} in \texttt{REBOUND} \citep{ReinHu12}. \texttt{IAS15} assumes an adaptive integration timestep that is automatically adjusted to increase the sensitivity of the simulation at close encounters with the star or planets \citep{ReinSpiegel15}.\\

The goal of our simulations is to track the orbital evolution of particles and determine if they may cross the gap that separates the outer disc from the inner disc, and evolve onto inclined, eccentric orbits. We define the inner edge of the gap as the outer boundary of the inner disc, at roughly 18 au from the star \citep{Benisty+21}. We note that further N-body simulations that track the dynamics of these particles in the inner regions of the system are currently unreliable due to the high uncertainty of the constituents of the system within approximately 10 au: the existence of any additional inner planets could dramatically change the nature of the dynamical evolution of particles in the inner system, exciting the inclination distribution of scattered planetesimals even further. Our simulations include gravitational forces only, and assume gas to be effectively absent, neglecting any gas drag force in our particles. This approach suffices in the gap between the inner and outer disc, where gas is observed to be depleted \citep{Keppler+19, Facchini+21}. In the outer disc, the timescale for a planetesimal of 0.3 km or larger to decelerate due to gas drag is significantly higher than the age of PDS 70 \citep{Muller+18} (see \citep{Eriksson+21} for details on drag force), justifying our assumption to neglect gas drag in favour of pure N-body interactions. Within the inner disc, the gas drag timescale is comparable to the age of the system. However, we only probe the orbital parameters of particles until the moment that they have ``crossed the gap'', without carrying on the simulation in the inner system.\\

\begin{figure}[t]
\centering
\includegraphics[width=\textwidth]{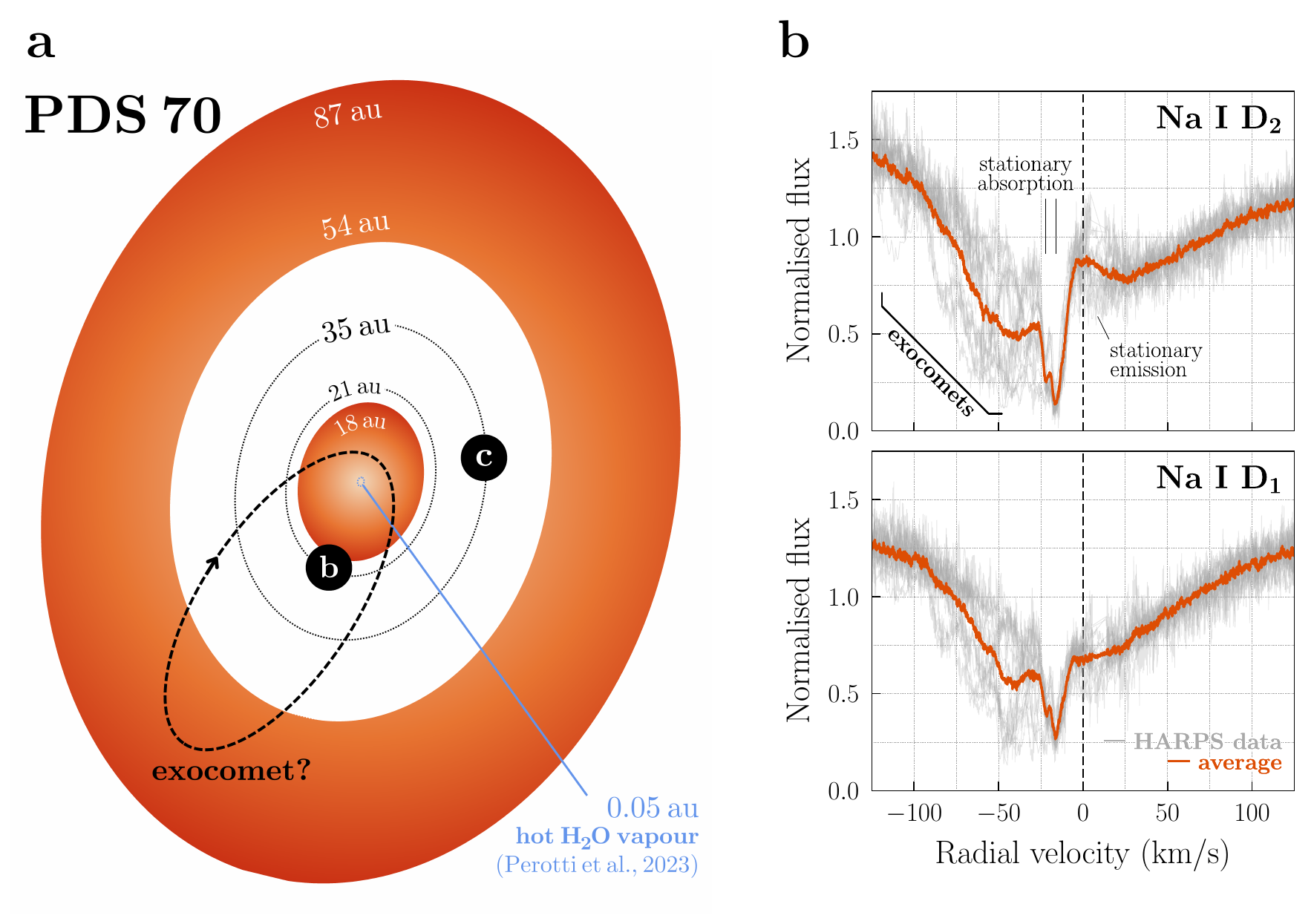}
\caption{\textbf{Schematic of the PDS 70 system and variability in the sodium lines of PDS 70.} \textbf{a}, Schematic of PDS 70, containing PDS 70 b and PDS 70 c in black, and the inner and outer discs in orange. The interior 0.05 au of the disc (blue) is observed to contain \ch{H2O} vapour \citep{Perotti+23}, which we propose is caused by sublimating exocomets coming from the outer disc. Disc sizes and orbital distances are to scale. Planet sizes are increased 200$\times$ for better visualisation. \textbf{b}, HARPS data around the Na\,I D lines (top: 588.996 nm and bottom: 589.593 nm) at 18 different epochs between March 29 and May 13, 2018. Solid grey lines indicate the combined spectra of each epoch, for which a different number of exposures were taken (see Table \ref{table1} and Methods, subsection \nameref{sec:observations_analysis_methods}). In each spectra we have identified two blue-shifted stationary absorption lines, a stationary emission line, and additional 1 to 4 variable absorption line components, all superimposed on the broad stellar photospheric absorption line. Orange lines represent the average between all 18 spectra. Spectra are centred (dashed line) at a systemic velocity of 6.01 km/s, derived from the data (see Methods, subsection \nameref{sec:vsys_methods}). Source data are provided (see \nameref{sec:data_availability}).}
\label{fig1}
\end{figure}

We simulate a total of 10\,000 massless test particles randomly distributed across a disc spanning from 54 to 87 au, corresponding to the outer disc boundaries of PDS 70 \citep{Benisty+21} (see Fig. \ref{fig1}a). Particles are initialised with zero eccentricity and random inclinations between $\pm \, 10^{\circ}$, motivated by the spread of inclinations observed in the present-day Kuiper belt \citep{Gulbis+10}. Here we note that we also tested lower inclinations closer to the disc mid-plane, which yielded similar results. This is likely because the planets are not perfectly coplanar \citep{Trevascus+25}, and even a small mutual inclination between the planets is sufficient to excite particles vertically. Following similar N-body simulations recently carried out by \citep{Trevascus+25}, we perform two sets of simulations: the first assuming the presence of confirmed planets PDS 70 b and c only, and the second also assuming planet candidate PDS 70 d \citep{Mesa+19, Christiaens+24}. Assuming stability of the planetary orbits, we used the median value of the posterior of the ``stable (including N-body)'' orbital parameters from Tables 3 and 4 of \citep{Trevascus+25}, also listed in Supplementary Table \ref{suptable5}, for the first (two-planet) and second (three-planet) set of simulations: mass ($M$), semi-major axis ($a$), eccentricity ($e$), inclination ($i$), argument of periastron ($\omega$), and longitude of the ascending node ($\Omega$). We note that this results in different properties for the two sets of simulations, in particular in the planetary masses. In the two-planet case, the masses of planets b and c are assumed to be 1.4 and 6.4 M$_{\rm Jup}$, respectively, while 0.7, 2.4, and 0.4 M$_{\rm Jup}$ are assumed for planets b, c, and d in the three-planet case \citep{Trevascus+25}. To reduce computational time, the code parallelises 50 simulations, each tracking the star, the two or three planets, and 200 of the 10\,000 test particles. This parallelisation is possible because the particles are massless and do not affect each other gravitationally. In each simulation, the planets are initialised with a phase taken from a random distribution from $0$ to $2\pi$. We integrate for 5 Myr, corresponding to the estimated age of the system \citep{Muller+18}, with a maximal timestep of 0.2 yr.\\

It is known that perturbations between the planets can lead to instability \citep{Trevascus+25}, especially if the planets have masses higher than several Jupiter masses. Generally, a fraction of these simulations are unstable over 5 Myr. We define a system to be stable when it satisfies the following criteria: i. all planets are bound to the system, i.e. $e < 1$; ii. planets do not cross orbits, i.e. $a_{\rm b}(1+e_{\rm b}) < a_{\rm c}(1-e_{\rm c})$, and $a_{\rm d}(1+e_{\rm d}) < a_{\rm b}(1-e_{\rm b})$ if PDS 70 d is included, following \citep{Wang+20}; and iii. neither the outermost planet, nor consequently, any of the others, cross the outer edge of the gap, i.e. $a_{\rm c}(1+e_{\rm c}) \leq 54$ au, similar to \citep{Trevascus+25}. 1 out of the 50 simulations in the first set of simulations (i.e. including planets b and c only) did not satisfy these stability criteria, and was therefore disregarded from further analysis, bringing the total number of particles to 9\,800. When planet d was included, 8 out of the 50 simulations did not satisfy these criteria, bringing the total number of particles to 8\,400 in the second set of simulations. We track the orbital evolution of these particles to determine if at any time during the simulation their periastron distance is equal to or less than 18 au from the star \citep{Benisty+21}, or if they are ejected from the system. Fig. \ref{fig8} shows the relative number of particles that we classify as crossing the gap or ejected for the two sets of simulations, as well as the orbital parameters (periastron distance, inclination, semi-major axis, and eccentricity) of the scattered particles at the time they cross the gap.\\

In addition to the findings presented in \nameref{sec:exocomet_scattering} and \nameref{sec:exocomet_inclinations_eccentricities}, we note that particles that initialise closer to the inner boundary of the outer disc, and consequently closer to PDS 70 c, are more subject to be scattered inwards due to gravitational interactions with PDS 70 c (Fig. \ref{fig8}b and Fig. \ref{fig8}f). This is true for both the two-planet and the three-planet cases, where PDS 70 c is located at semi-major axes of 33.9 and 35.3 au, respectively \citep{Trevascus+25}. However, in both cases, particles initialised at large distances from PDS 70 c are still efficiently scattered, which suggests that low-order mean-motion resonances are a likely driver for the scattering of particles in the outer disc (Fig. \ref{fig8}c and Fig. \ref{fig8}g). We note that, in both cases, the initial positions of particles that cross the gap reveal regions where particles are scattered, but their periastron distances never cross the inner disc during the simulation, indicating they remain in the outer disc. Furthermore, we note that particles reach a wide range of eccentricities and inclinations regardless of where in the outer disc they start the simulation (Fig. \ref{fig8}d and Fig. \ref{fig8}h). However, the highest inclinations and eccentricities are achieved by particles that started the simulation closer to the inner boundary of the outer disc (approximately 54--75 au), i.e. closer to PDS 70 c, in both the two- and three-planet cases (Fig. \ref{fig8}c and Fig. \ref{fig8}g). We conclude that the inferred fraction of particles that cross the gap, as well as their eccentricity and inclination are dependent on their initial semi-major axes, but are mainly dependent on the mass of PDS 70 c. Even so, both the two-planet (higher planet c mass) and the three-planet (lower planet c mass) cases produce highly inclined, eccentric particles with periastron distances below 18 au.\\

\subsection*{Periastron distances of transiting exocomets}

An outstanding question is where in the system these exocomets are located as they become spectroscopically visible during their transit across the star. Under the assumption that the cloud has the same velocity as the evaporating nucleus that is on a bound Keplerian orbit \citep[e.g.][]{Kiefer+14}, the observed radial velocity provides a maximum distance (as high velocities are only possible for small orbital distances and vice versa). Kiefer et al. \citep{Kiefer+14} uses a formalism to constrain the instantaneous distance further, by assuming that the eccentricities are high, such that a parabolic approximation can be made. Under such assumptions and given an observed radial velocity, the instantaneous periastron distance is only degenerate with the longitude of periastron. Fig. \ref{fig9} shows the expected range of radial velocities of transiting exocomets as a function of orbital distance and longitude of periastron. For comets that are blue-shifted (i.e. receding from periastron), the longitude of periastron is necessarily between 180$\degree$ and 360$\degree$ (see equation 22 of \citep{Kiefer+14}, and note that in this definition of longitude of periastron in \citep{Kiefer+14}, 180$\degree$ corresponds to the periastron being located directly behind the star, and 0 or 360$\degree$ corresponds to the periastron being located between the star and the observer). This analysis suggests that a wide range of orbital distances are possible, up to about 4 au for the slowest exocomets detected, while the fastest comets require distances less than approximately 0.1 au.\\

Further constraints on the orbital distance can be imposed by regarding the temperature required to sublimate sodium. The equilibrium sublimation temperature of Na\,I is 958 K, calculated for a solar-composition gas at $10^{-4}$ bar pressure \citep{Lodders03}. We take here this equilibrium temperature to represent the approximate temperature around which sodium atoms within the planetesimal’s silicate minerals begin to undergo rapid sublimation. This corresponds to the temperature at a distance of 0.05 au from the star, which suggests that these exocomets must be located within approximately 0.05 au or closer at the time that they are observed spectroscopically. Interestingly, an orbital distance of 0.05 au coincides with the extent of the emitting disc of water molecules observed using JWST/MIRI \citep{Perotti+23}. However, there are several theoretical objections to this quantification of the sublimation process. First, the sublimation temperature of 958 K by \citep{Lodders03} is not applicable for low-gravity exocomets in a vacuum that have an unknown composition. Secondly, release of sodium from rocky material can occur even at much greater distances from the star. Clear examples are offered by bodies in the Solar System, such as sublimated sodium tails observed in comets, far beyond the sodium sublimation line at 0.08 au from the Sun (e.g. comet McNaught C/2006 P1, observed at 0.2 au \citep{Leblanc+08}, and comet Hale-Bopp, observed at 0.98 au \citep{Cremonese+97}). Mercury also exhibits an extended tail of sodium, liberated from the surface by sputtering at a semi-major axis of 0.39 au \citep{McGrath+86}. For these reasons, we are not yet able to further constrain where these comets are located when viewed in transit, and more sophisticated simulations of the sublimation of sodium are necessary to achieve this.\\

\clearpage

\begin{table*}
\centering
\caption{List of HARPS observations. SNR is the average of the signal-to-noise ratio in order 56 (around the Na\,I D lines) of all spectra in each night.}
\vspace{0.5cm}
\resizebox{0.9\textwidth}{!}{
\begin{tabular}{lccccc}
\hline
Programme ID/Run & PI & \begin{tabular}[c]{@{}c@{}}Date\\ (year-month-day)\end{tabular} & \begin{tabular}[c]{@{}c@{}}Spectra\\ per night\end{tabular} & \begin{tabular}[c]{@{}c@{}}Exposure time\\ per spectrum (s)\end{tabular} & SNR \\ \hline
\multirow{13}{*}{098.C-0739(A)} & \multirow{13}{*}{A.-M. Lagrange} & 2018-03-29 & 6 & 900 & 22.67 \\
 &  & 2018-03-30 & 6 & 900 & 15.83 \\
 &  & 2018-03-31 & 6 & 900 & 27.67 \\
 &  & 2018-04-18 & 1 & 1800 & 30.50 \\
 &  & 2018-04-19 & 2 & 900 & 11.70 \\
 &  & 2018-04-20 & 2 & 1800 & 33.25 \\
 &  & 2018-04-21 & 1 & 1800 & 31.00 \\
 &  & 2018-04-22 & 1 & 1800 & 33.10 \\
 &  & 2018-04-23 & 1 & 1800 & 30.90 \\
 &  & 2018-05-01 & 2 & 1800 & 21.00 \\
 &  & 2018-05-06 & 2 & 1800 & 36.85 \\
 &  & 2019-02-13 & 2 & 1800 & 16.60 \\
 &  & 2019-05-01 & 2 & 2400 & 32.30 \\ \hline
1101.C-0557(A) & A.-M. Lagrange & 2018-05-13 & 2 & 1800 & 26.25 \\ \hline
\multirow{3}{*}{101.C-0557(D)} & \multirow{3}{*}{A.-M. Lagrange} & 2020-02-25 & 2 & 1800 & 24.55 \\
 &  & 2020-02-26 & 2 & 1800 & 35.60 \\
 &  & 2020-02-27 & 2 & 1800 & 32.90 \\ \hline
\multirow{2}{*}{1101.C-0557(D)} & \multirow{2}{*}{A.-M. Lagrange} & 2020-02-29 & 2 & 1800 & 33.05 \\
 &  & 2020-03-12 & 2 & 1800 & 28.95 \\ \hline
\multirow{3}{*}{0104.C-0418(C)} & \multirow{3}{*}{A.-M. Lagrange} & 2020-03-13 & 2 & 1800 & 36.30 \\
 &  & 2020-03-14 & 2 & 1800 & 31.05 \\
 &  & 2020-03-15 & 2 & 1800 & 30.25 \\ \hline
\end{tabular}
}
\label{table1}
\end{table*}

\clearpage

\clearpage 

\begin{figure}[t]
\centering
\includegraphics[width=\textwidth]{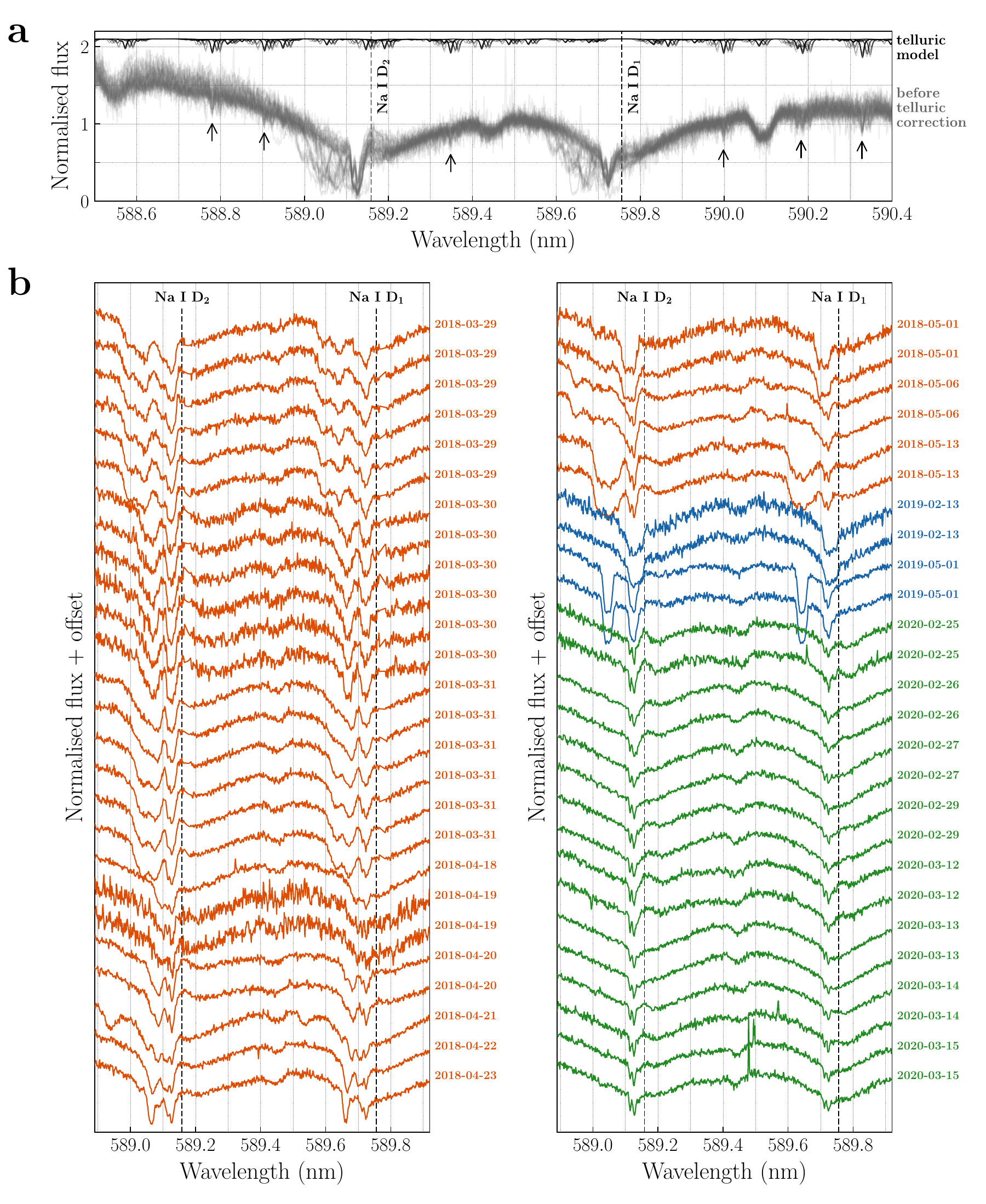}
\caption{\textbf{Collection of PDS 70 HARPS data around the sodium lines.} \textbf{a}, Collection of 52 HARPS spectra of PDS 70 around the Na\,I D lines (dashed black lines, 588.996 nm and 589.593 nm), taken between 2018 and 2020. Telluric models for all spectra are shown at the top (black lines; see Methods, subsection \nameref{sec:observations_analysis_methods}), and arrows show the position of the strongest telluric lines. \textbf{b}, A detailed view of each exposure after telluric correction. Orange, blue, and green lines represent spectra taken in 2018, 2019, and 2020, respectively (see Table \ref{table1} for Programme IDs and dates). Source data are provided (see \nameref{sec:data_availability}).}
\label{fig2}
\end{figure}

\clearpage 

\begin{figure}[t]
\centering
\includegraphics[width=\textwidth]{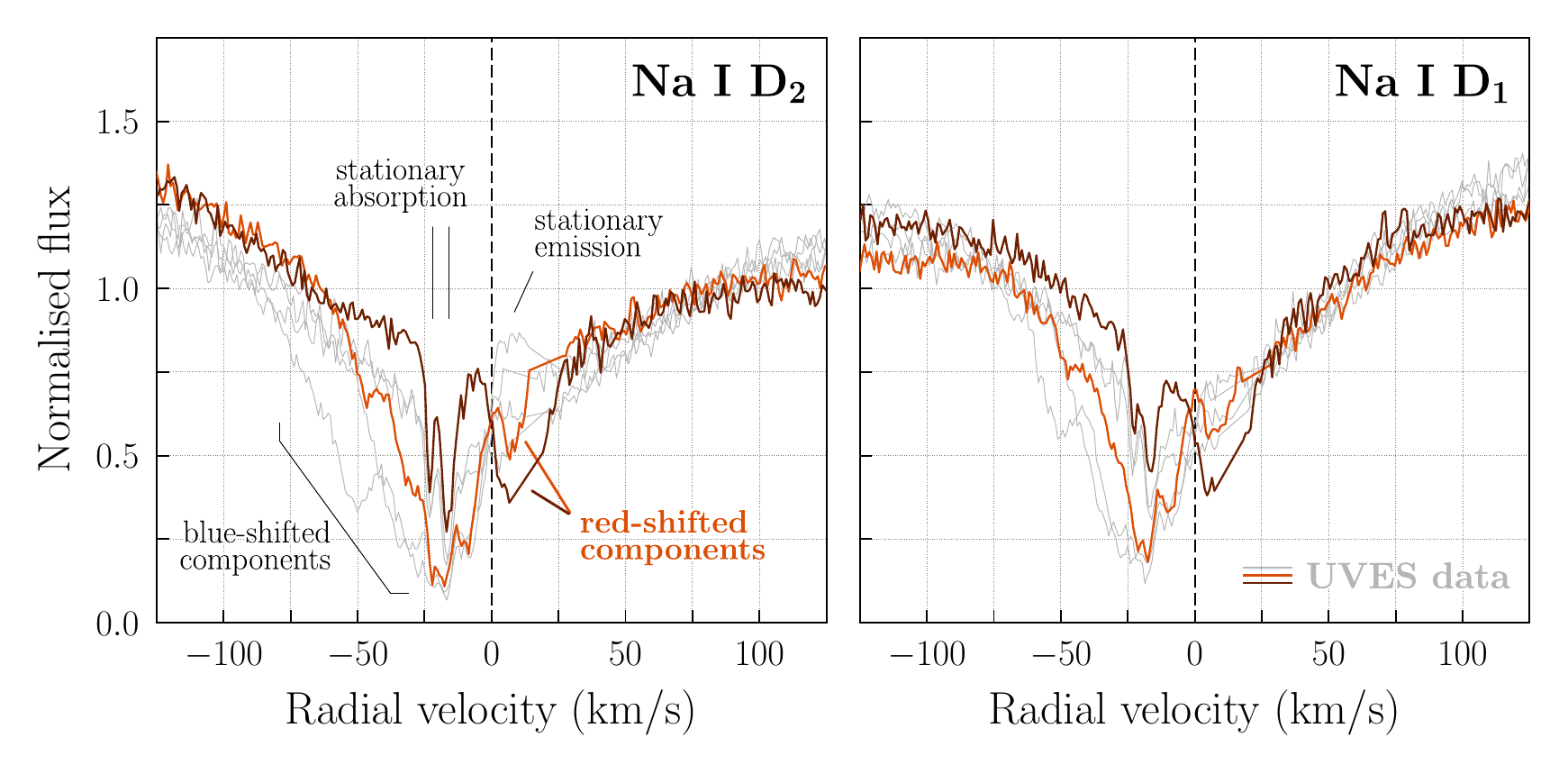}
\caption{\textbf{Variability in the sodium lines of PDS 70 in UVES data, including red-shifted lines.} UVES data around the Na\,I D lines at 6 different epochs: 2026-01-16, 2026-01-18, 2026-01-19, 2026-03-18, 2026-03-22, and 2026-03-25 (Programme ID 0116.C-0329, PI A. Novais). In each epoch we have identified variable blue-shifted absorption lines, as well as the same stationary emission line approximately at the stellar rest frame, and stationary absorption lines at approximately $-22$ and $-16$ km/s, also present in the 2018 HARPS data (see Fig. \ref{fig1}b). In addition, variable red-shifted absorption lines are present in two epochs: 2026-01-18 (represented in orange) and 2026-03-22 (represented in brown). Spectra are centred (dashed line) at a systemic velocity of 6.01 km/s, derived from the HARPS data (see Methods, subsection \nameref{sec:vsys_methods}).}
\label{fig3}
\end{figure}

\clearpage 

\begin{figure}[t]
\centering
\includegraphics[width=\textwidth]{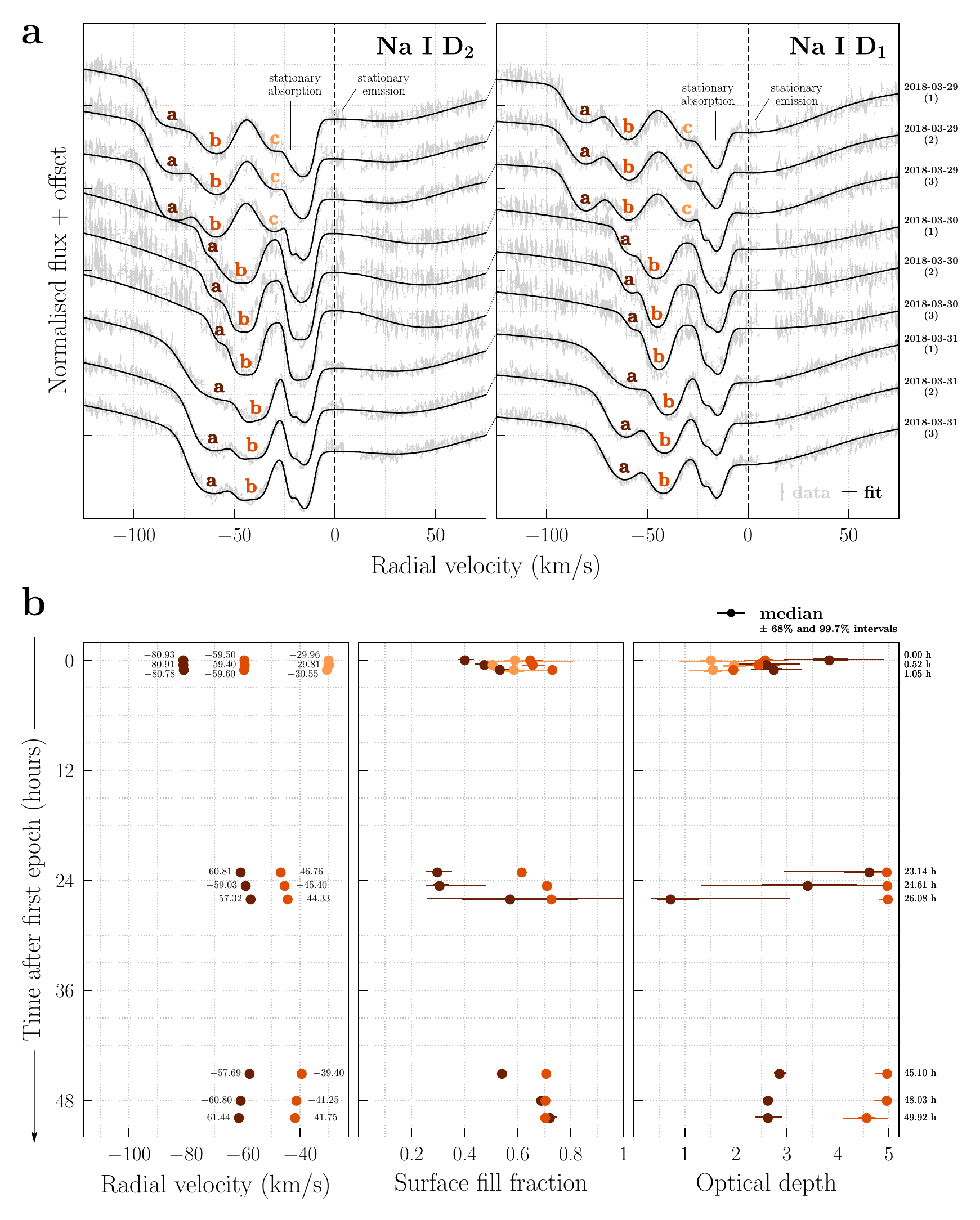}
\caption{\textbf{Variability in the sodium lines of PDS 70 within 50 hours.} \textbf{a}, HARPS spectra around the Na\,I D lines at nine different epochs obtained within 50 hours, in March 29, 30, and 31, 2018. Circles with error bars represent the exposures in each epoch (see Table \ref{table1} and Methods, subsection \nameref{sec:observations_analysis_methods}), and solid lines indicate the best-fit spectra for the exposures combined. Variable Na\,I absorption line components are indicated from most blue-shifted (a) to least blue-shifted (c). Spectra are centred (dashed line) at a systemic velocity of 6.01 km/s, derived from the data (see Methods, subsection \nameref{sec:vsys_methods}). \textbf{b}, Parameters fitted from the HARPS data as a function of time after the first epoch (2018-03-29 (1)). Markers, thick, and thin error bars represent the median and the 68\% and 99.7\% confidence intervals, corresponding to the values between the 16--84 percentiles, and the 0.135--99.865 percentiles of the posterior distributions, respectively. Posteriors suggest the Na\,I gas in front of the star is optically thick ($> 1$), at high blue-shifted velocities, and with surface fill fractions that are less than the size of the projected stellar disc ($< 1$), indicating that the sodium clouds are confined. Source data are provided (see \nameref{sec:data_availability}).}
\label{fig4}
\end{figure}

\clearpage

\begin{figure}[t]
\centering
\includegraphics[width=\textwidth]{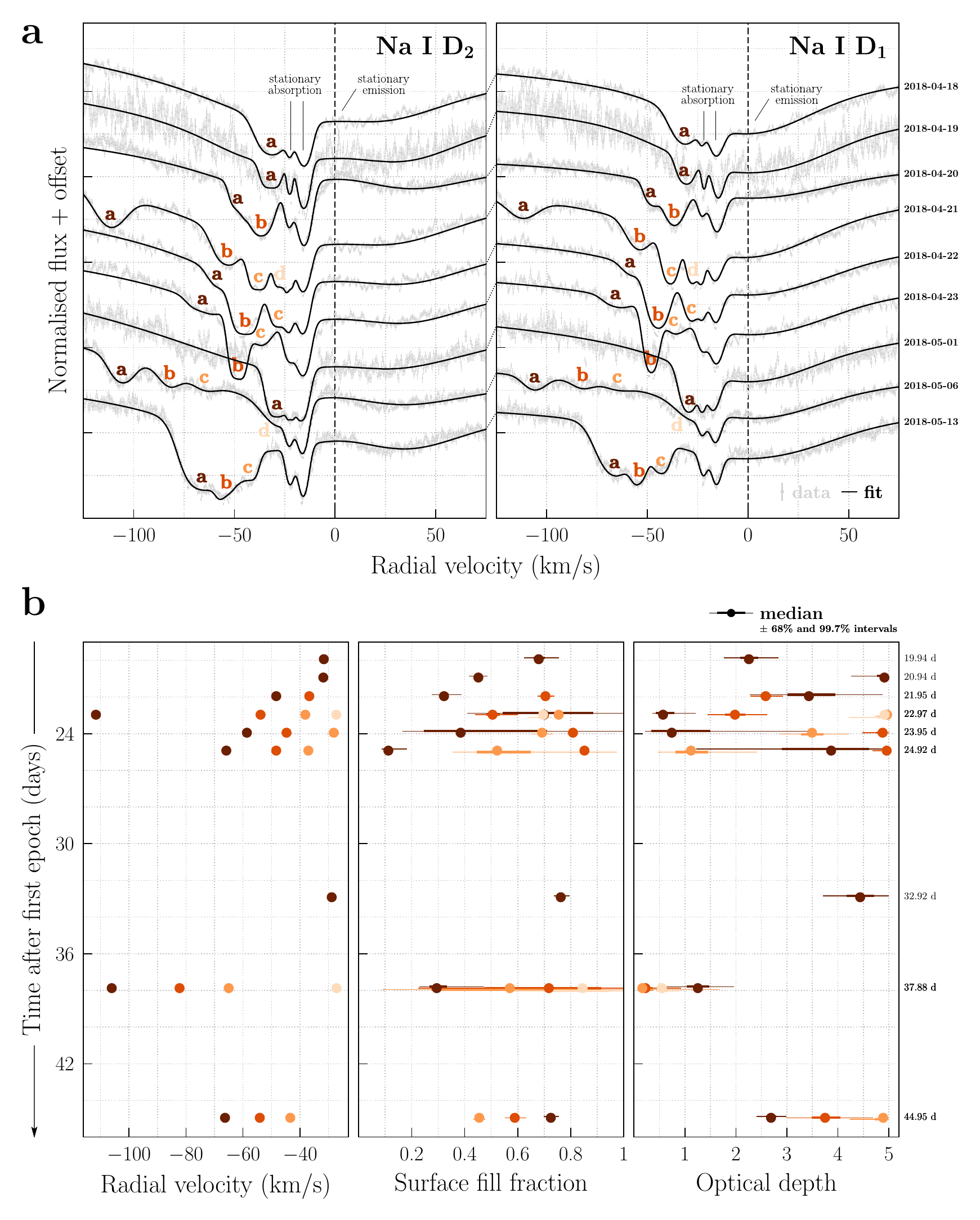}
\caption{\textbf{Variability in the sodium lines PDS 70 within 25 days.} \textbf{a}, HARPS spectra around the Na\,I D lines for the remaining epochs not present in Fig. \ref{fig4} (April 18 to May 13, 2018). Circles with error bars represent the exposures in each epoch (see Table \ref{table1} and Methods, subsection \nameref{sec:observations_analysis_methods}), and solid lines indicate the best-fit spectra for the exposures combined. Variable Na\,I absorption line components are indicated from most blue-shifted (a) to least blue-shifted (d). Spectra are centred (dashed line) at a systemic velocity of 6.01 km/s, derived from the data (see Methods, subsection \nameref{sec:vsys_methods}). \textbf{b}, Parameters fitted from the HARPS data as a function of time after the first epoch of Fig. \ref{fig4} (2018-03-29 (1)). Markers, thick, and thin error bars represent the median and the 68\% and 99.7\% confidence intervals, corresponding to the values between the 16--84 percentiles, and the 0.135--99.865 percentiles of the posterior distributions, respectively. Source data are provided (see \nameref{sec:data_availability}).}
\label{fig5}
\end{figure}

\clearpage

\begin{figure}[t]
\centering
\includegraphics[width=\textwidth]{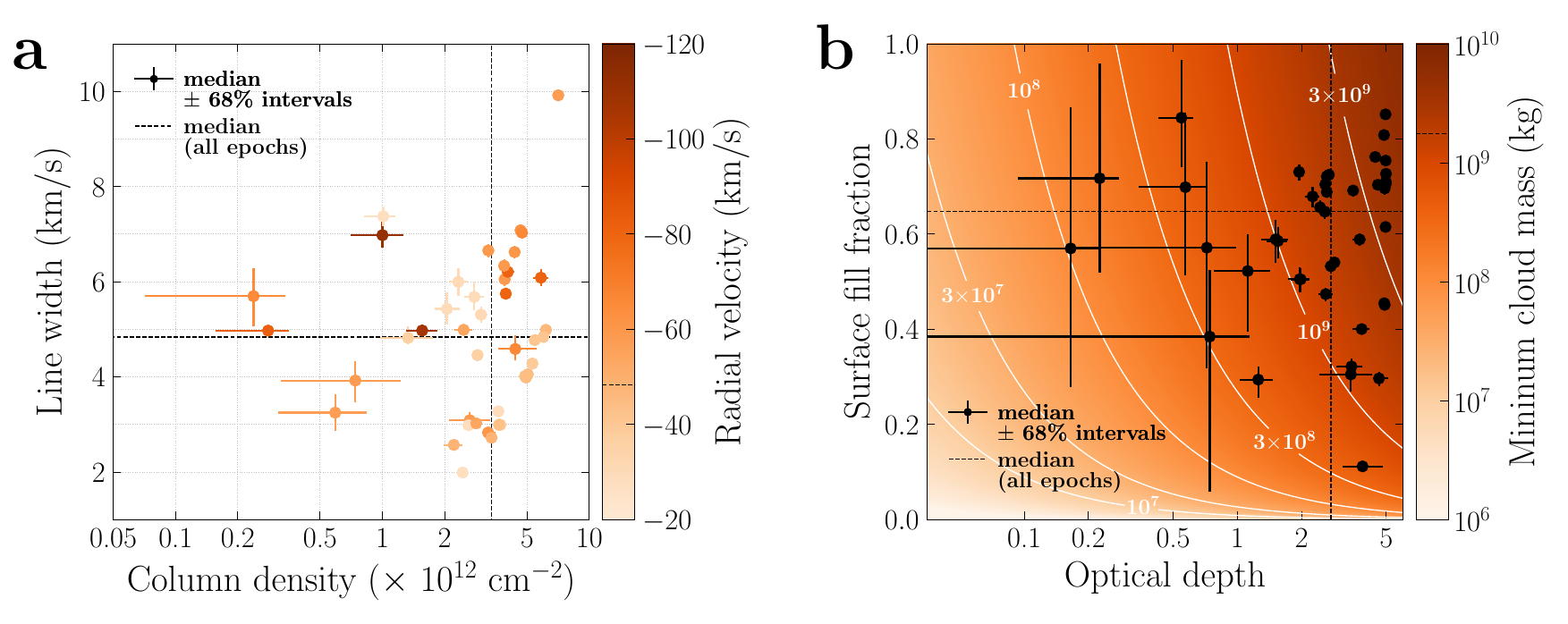}
\caption{\textbf{Properties of the PDS 70 variable sodium absorption lines.} Circles represent median values fitted from the 43 variable Na\,I absorption lines found in PDS 70 HARPS spectra from March 29 to May 13, 2018. Error bars represent the 68\% confidence interval (between the 16--84 percentiles) of the posterior distributions. Dashed lines represent the median best-fit values across all epochs. \textbf{a}, Gaussian line width as a function of column density and radial velocity of the variable Na\,I absorption lines. Column densities are calculated from radial velocity, Gaussian line width, and optical depth of all fit samples (see Methods, subsection \nameref{sec:column_density_methods}). Our measurements suggest sodium column densities between $2.4 \times 10^{11}$ and $7.1 \times 10^{12}$ cm$^{-2}$, with a median value of $3.4 \times 10^{12}$ cm$^{-2}$. \textbf{b}, Surface fill fraction as a function of optical depth using a slab model, in order to estimate a lower limit of the mass of these sodium clouds. Colour bar represents mass values for the median line width of 4.8 km/s. We estimate the minimum mass of each of these sodium clouds to be between $1.3 \times 10^{8}$ and $3.9 \times 10^{9}$ kg, with a median value of $1.8 \times 10^{9}$ kg. Source data are provided (see \nameref{sec:data_availability}).}
\label{fig6}
\end{figure}

\clearpage

\begin{figure}[t]
\centering
\includegraphics[width=\textwidth]{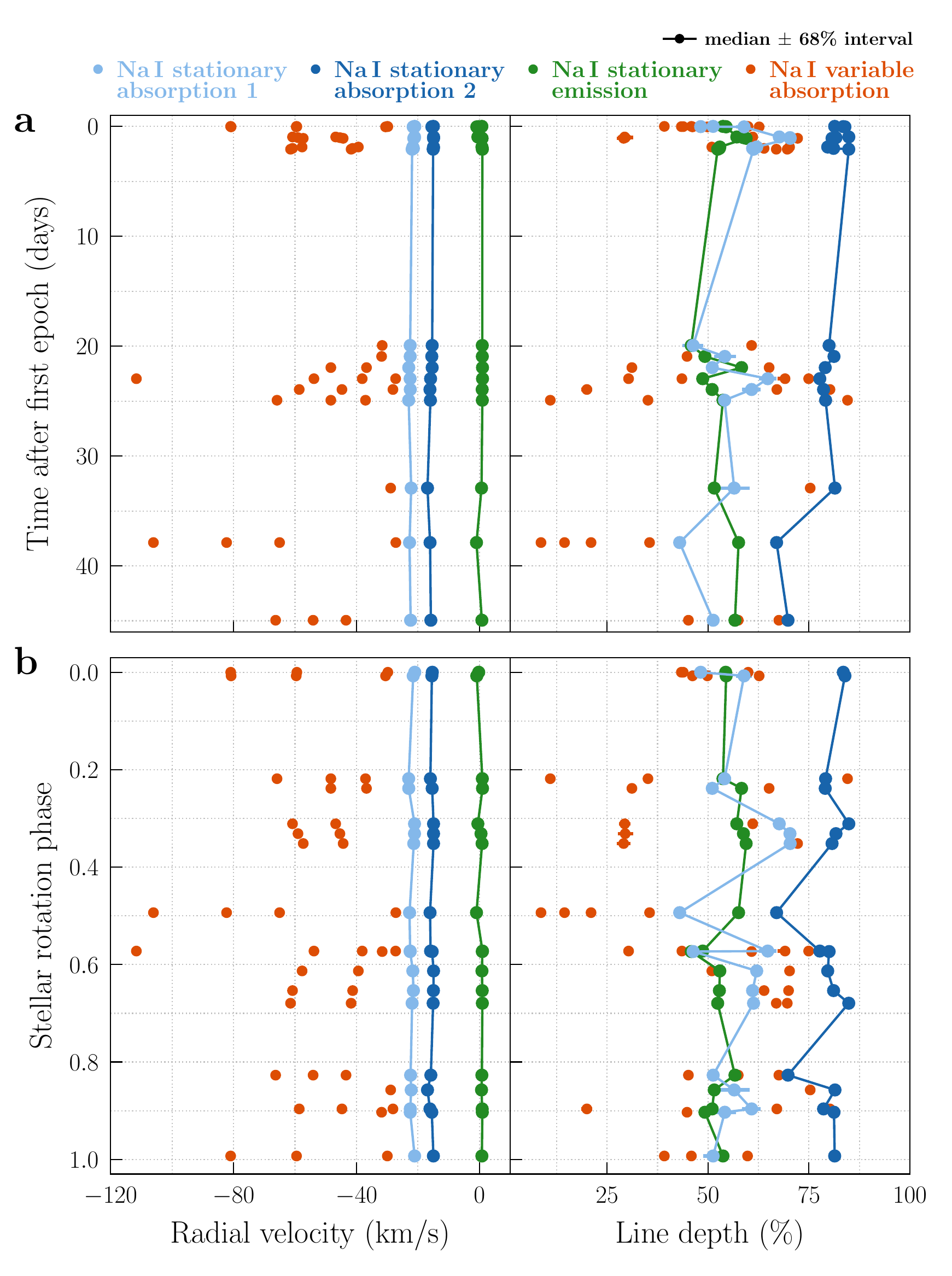}
\caption{\textbf{Stationary sodium line components as a function of time and stellar rotation period}. Radial velocity and line depth measurements at line centre of the sodium line components observed in 2018. Circles correspond to median values of the posterior distributions, while error bars represent the 68\% confidence interval (between the 16--84 percentiles). \textbf{a,} Measurements as a function of time after the first epoch (2018-03-29 (1)). \textbf{b,} Same as a, but as a function of the stellar rotation phase, assuming a rotation period of 3.03 days \citep{Thanathibodee+20}. Phase = 0 was taken at MJD (Modified Julian Date) = 0. The figure illustrates the two partially blended absorption components, stationary at roughly $-22$ and $-16$ km/s (light and dark blue, respectively), and the emission component, which is stationary at the stellar rest frame (green). Source data are provided (see \nameref{sec:data_availability}).}
\label{fig7}
\end{figure}

\clearpage

\begin{figure}[t]
\centering
\includegraphics[width=\textwidth]{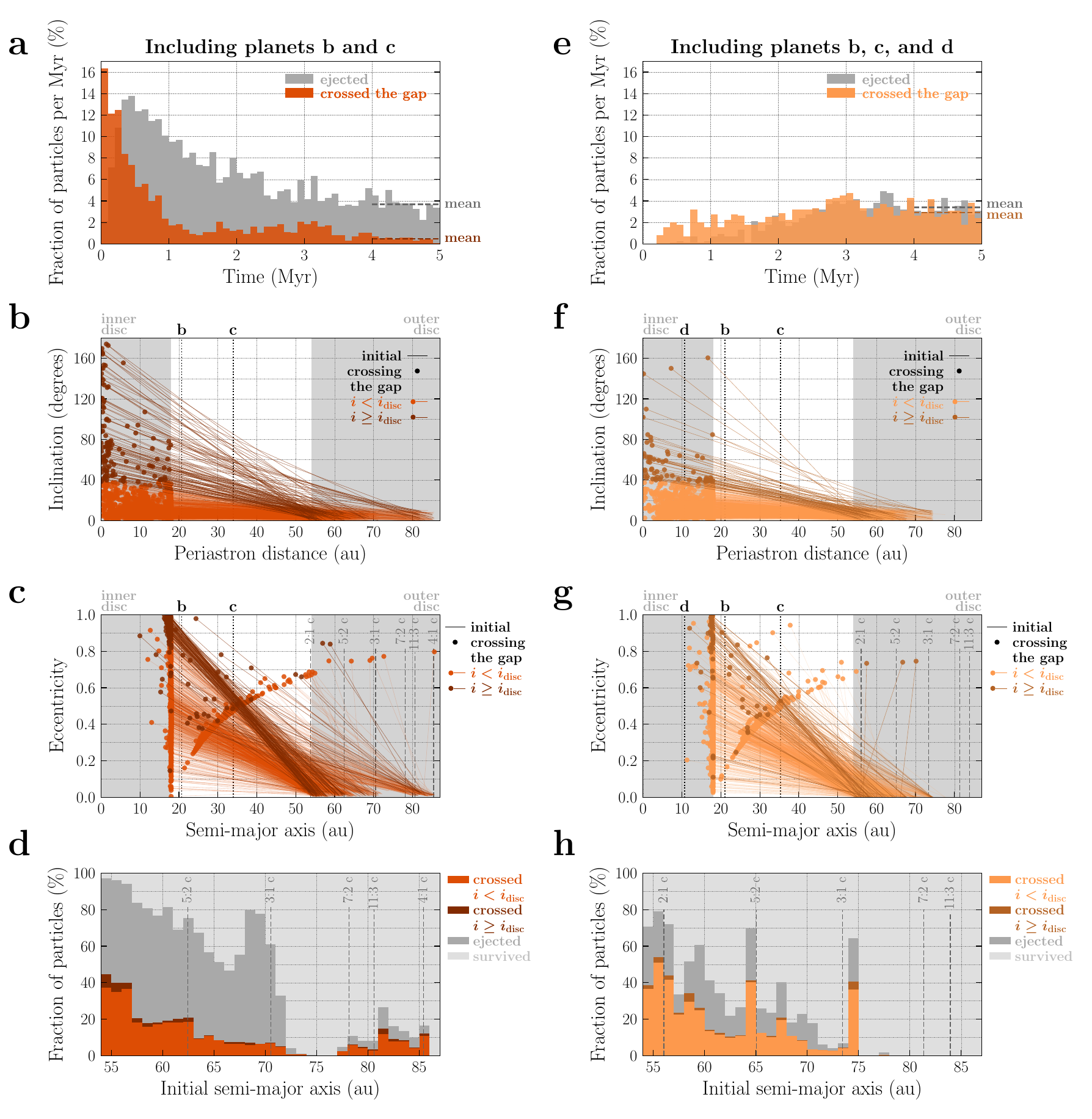}
\caption{\textbf{Orbital evolution of particles towards the inner disc from N-body simulations.} \textbf{a}, Relative number of particles that crossed the gap towards the inner disc (orange) or were ejected from the system (grey) as a function of time, from N-body simulations including PDS 70 b and c, out of a total of 9\,800 test particles. Here, ``crossed the gap'' means that the periastron distance of a particle is equal to or less than 18 au \citep{Benisty+21} from the star at any time during the simulation. Dashed lines indicate the mean fraction of ejected particles and particles that cross the gap over the last 1 Myr. Simulations ran for 5 Myr (see Methods, subsection \nameref{sec:nbody_simulations_methods}). \textbf{b}, Orbital inclination (relative to the plane of the disc) of the particles that crossed the gap as a function of their periastron distance. Lines show the initial parameters, and markers represent parameters measured at the time each particle crosses the gap. Darker shade highlights particles that achieve inclinations equal to or above the disc inclination of 38.3$\degree$ relative to the line of sight \citep{Keppler+19}. \textbf{c}, Orbital eccentricity of the particles that crossed the gap as a function of their semi-major axis. Colours are the same as in b. The location of mean-motion resonances with PDS 70 c are illustrated with dashed grey lines (2:1 c, 5:2 c, etc). \textbf{d}, The fate of particles as a function of where they were initialised in the outer disc. Particles that survived in the disc, i.e. particles with periastron distances that did not cross the gap and were not ejected, are represented as ``survived''. \textbf{e}, \textbf{f}, \textbf{g}, and \textbf{h}, Same as a, b, c, and d, but including planet candidate PDS 70 d, out of a total of 8\,400 particles. Source data are provided (see \nameref{sec:data_availability}).}
\label{fig8}
\end{figure}

\clearpage

\begin{figure}[t]
\centering
\includegraphics[width=\textwidth]{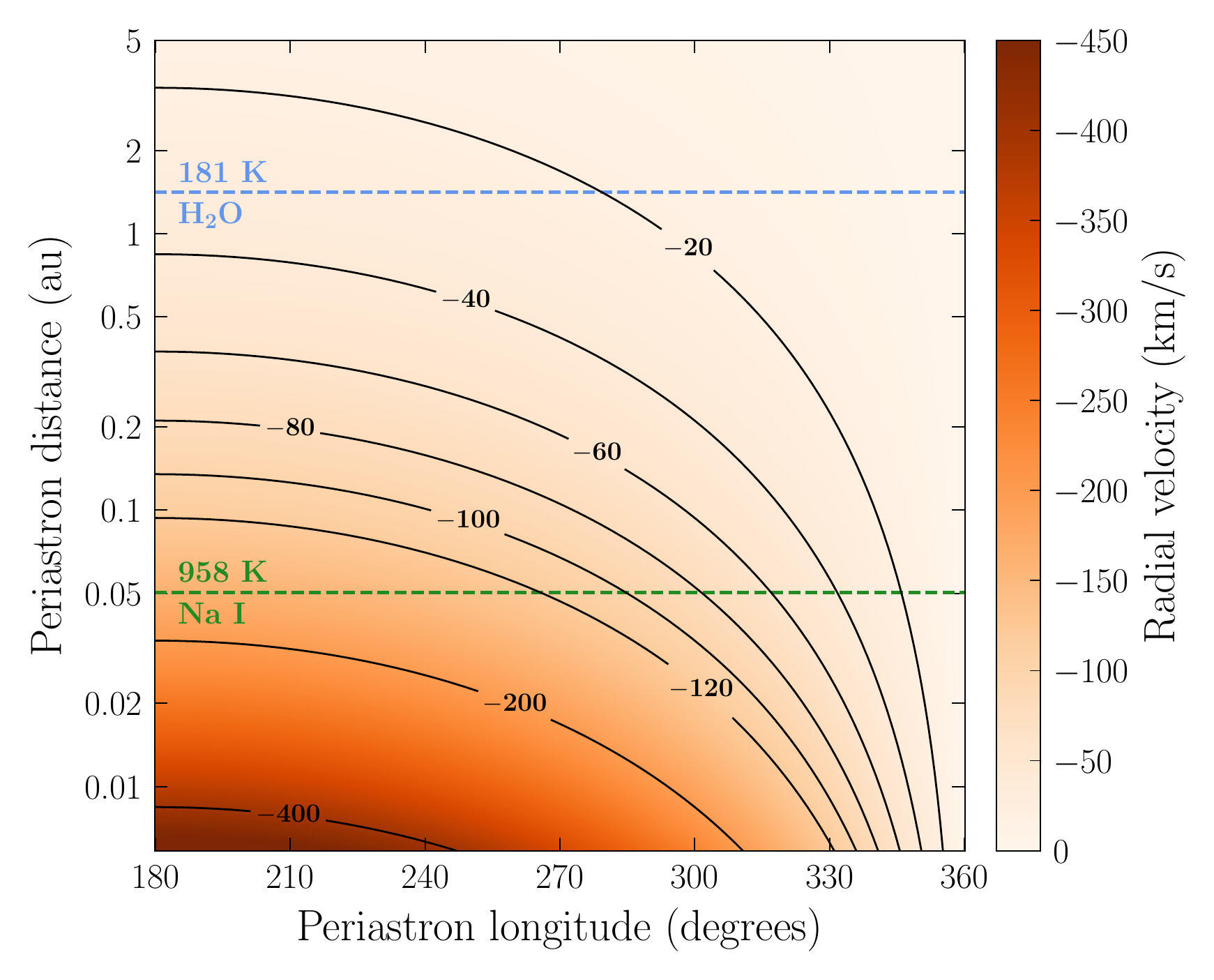}
\caption{\textbf{Radial velocities relative to the periastron distance and longitude in PDS 70.} Radial velocity values as a function of the periastron distance and the periastron longitude, following equation 22 of \citep{Kiefer+14} assuming a near-parabolic orbit, a stellar effective temperature of 3972 K \citep{PecautMamajek16, Keppler+18}, radius of 1.26 R$_\odot$ \citep{Keppler+18}, and mass of 0.76 M$_\odot$ \citep{Muller+18}. A periastron longitude of 180$\degree$ corresponds to the periastron lying behind the star along the line of sight. Dashed green and blue lines refer to the condensation temperatures of Na\,I and \ch{H2O} based on Solar System values from \citep{Lodders03}, at 0.05 au and 1.41 au, respectively.}
\label{fig9}
\end{figure}

\vspace{0.8cm}

\subsection*{Data availability}
\label{sec:data_availability}
The HARPS data used in this work are part of the following European Southern Observatory (ESO) Programme IDs: 098.C-0739, 101.C-0557, 1101.C-0557, 0104.C-0418, PI A.-M. Lagrange. Processed (wavelength calibrated and reduced) archival HARPS data are publicly available at the ESO Archive database \href{https://archive.eso.org/eso/eso_archive_main.html}{https://archive.eso.org/eso/eso\_archive\_main.html} by including the Programme ID in the field ``Program ID'' in the pink menu under ``Target, Program, and Scheduling Information'' and selecting `HARPS/LaSilla' in the ``Spectroscopy'' column in the blue menu under ``Observing Information''. The UVES data presented in Fig. \ref{fig3} are part of ESO Programme ID 0116.C-0329, PI A. Novais. The UVES data are currently subject to the standard one-year proprietary period, and will become publicly available through the ESO Science Archive upon expiration of this period. Once available, the UVES data can be accessed in the same way as for HARPS, but selecting ``UVES/VLT'' in the ``Spectroscopy'' column. Telluric-corrected HARPS data presented in Fig. \ref{fig1} and Fig. \ref{fig2} are available at \href{https://doi.org/10.24433/CO.1461255.v1}{https://doi.org/10.24433/CO.1461255.v1} in the `data' folder. Posteriors and best-fit models of the the Na\,I doublet line components of all epochs, presented in Figs. \ref{fig4}--\ref{fig7}, are also available at \href{https://doi.org/10.24433/CO.1461255.v1}{https://doi.org/10.24433/CO.1461255.v1} under `code/utils/fits'. Input parameters for N-body simulations with \texttt{REBOUND} are available in Supplementary Table \ref{suptable5}, and simulation outputs presented in Fig. \ref{fig8} are available within the Source Data file. Source data for Figs. \ref{fig1}, \ref{fig2}, \ref{fig4}--\ref{fig8}, as well as for supplementary figures and tables, are available as within the Source Data file, as well as at \href{https://doi.org/10.5281/zenodo.21628743}{https://doi.org/10.5281/zenodo.21628743}.\\

\subsection*{Code availability}
The telluric absorption modelling tool \texttt{molecfit}, described by \citep{Smette+15, Kausch+15}, is publicly available at 
\href{https://eso.org/sci/software/pipelines/skytools/molecfit}{https://eso.org/sci/software/pipelines/skytools/molecfit}. The sodium doublet line fitting code used in this work is publicly available at \href{https://doi.org/10.24433/CO.1461255.v1}{https://doi.org/10.24433/CO.1461255.v1}. Spectral line fits makes use of \texttt{NumPyro}, a probabilistic programming for \texttt{Python} powered by \texttt{JAX}, motivated and described by \citep{Phan+19, Bingham+19}, and publicly available at \href{https://num.pyro.ai/en/latest}{https://num.pyro.ai/en/latest}. The calculation of the systemic velocity uses the nested-sampling package \texttt{PyMultiNest}, described by \citep{Buchner+14}, and publicly available at \href{https://johannesbuchner.github.io/PyMultiNest}{https://johannesbuchner.github.io/PyMultiNest}. The abundance and fraction of Na\,I used in \nameref{sec:discwinds_methods} were calculated using version C25.00 of \texttt{Cloudy}, last described by \citep{Gunasekera+25}, and publicly available at \href{https://gitlab.nublado.org/cloudy/cloudy/-/wikis/home}{https://gitlab.nublado.org/cloudy/cloudy/-/wikis/home}. N-body simulations were done using \texttt{REBOUND}, described by \citep{ReinHu12}, and publicly available at \href{https://rebound.readthedocs.io/en/latest}{https://rebound.readthedocs.io/en/latest}. Figure 1a was generated using Canva \href{https://www.canva.com}{https://www.canva.com}. All other figures were made with \texttt{Matplotlib} v.3.8.0, developed by \citep{Hunter07} under the licence at \href{https://matplotlib.org}{https://matplotlib.org}.\\

\subsection*{Funding statement}
A.N., H.J.H., and B.P. acknowledge financial support from The Fund of the Walter Gyllenberg Foundation. H.J.H. acknowledges support of grants from eSSENCE (grant number eSSENCE@LU 9:3), the Swedish National Research Council (project number 2023-05307), The Royal Physiographic Society, and The Crafoord foundation. A.J. acknowledges funding from the Carlsberg Foundation (Semper Ardens: Advance grant FIRSTATMO). A.S.M., K.J., and L.G. declare no relevant funding.\\

\subsection*{Author contributions}
A.N. participated in the development of the concept of this research, led the scientific discussion, performed the data analyses, and produced all the figures. H.J.H. participated in the development of the research concept, co-led the scientific discussion, and developed the code for fitting sodium doublet lines. A.N. and H.J.H. wrote the manuscript. A.S.M. performed N-body simulations, calculated the efficiency of particle scattering from the outer to the inner disc, and wrote the corresponding section. B.P. performed the derivation of the systemic velocity and wrote the corresponding section. A.J. provided insights on protoplanetary disc structures and water delivery theories in the Solar System. K.J. and H.J.H. provided insights on the study of exocomets on $\beta$ Pictoris and dynamical simulations. L.G. provided insights on stellar spectroscopy. All authors actively participated in the discussion of the results and commented on the manuscript.\\

\subsection*{Competing interests}
The authors declare no competing interests.\\

\clearpage

\section*{Supplementary Information}

\setcounter{figure}{0}
\renewcommand{\figurename}{Supplementary Fig.}
\setcounter{table}{0}
\renewcommand{\tablename}{Supplementary Table}

\begin{figure}[!h]
\centering
\includegraphics[width=\textwidth]{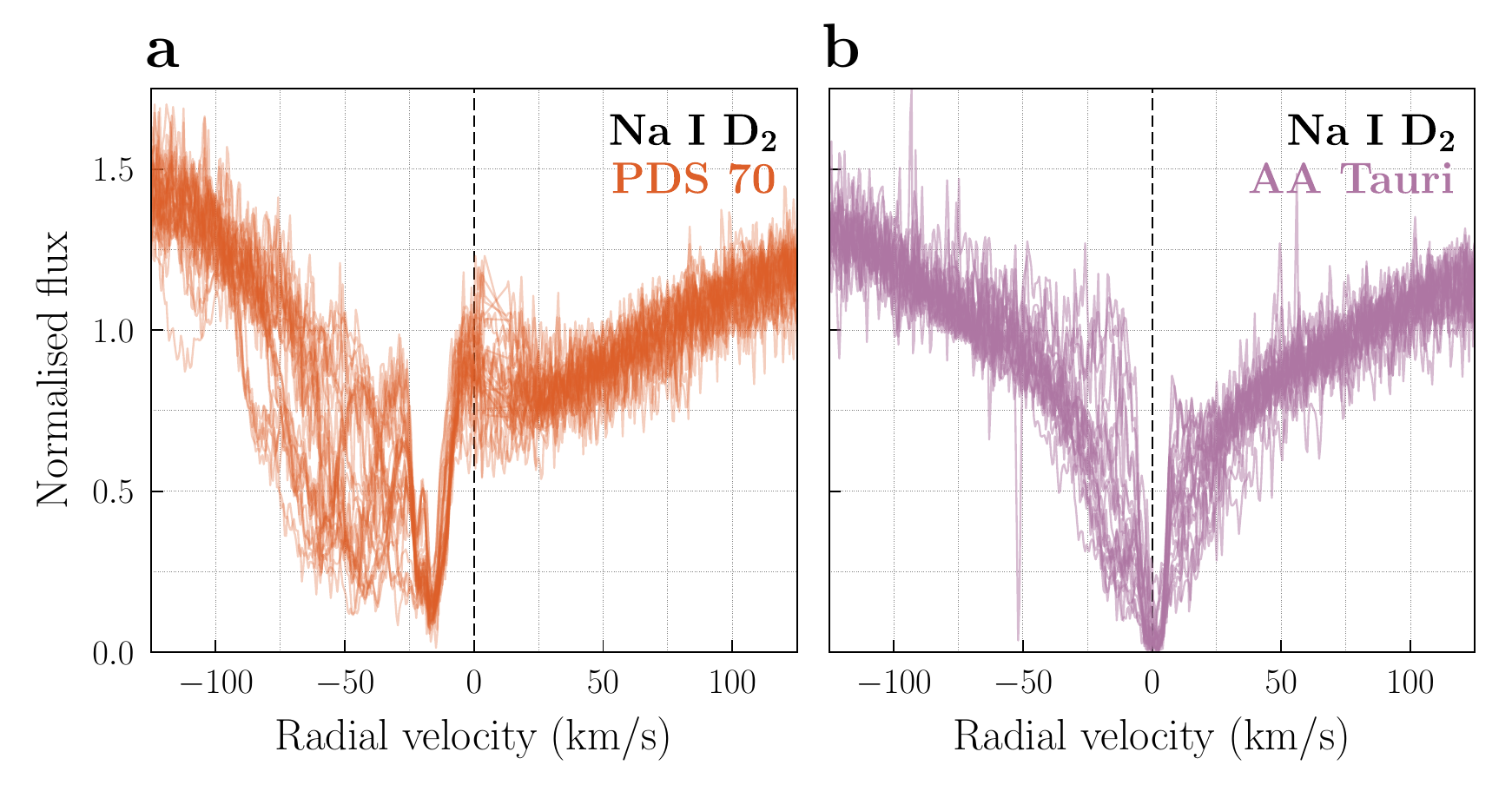}
\caption{\textbf{Sodium lines in \pds and AA Tauri.} Radial velocities derived from HARPS data around the Na\,I D$_2$ line (588.996 nm) of K7 T Tauri stars \pds (\textbf{a,} orange) and AA Tauri (\textbf{b,} purple). Spectra are centred (dashed line) at a system radial velocity of 6.01 km/s for \pds (see Methods, subsection Derivation of the systemic velocity) and 17.20 km/s for AA Tau \citep{Donati+10}. Data of \pds consist on 32 spectra taken on 12 nights between 2018-03-29 and 2018-05-13 (see Table \ref{table1} in the main text for Programme IDs). Data of AA Tau consist on 21 spectra taken between 2004-10-10 and 2004-12-02 (Programme ID 074.C-0221, PI Bouvier). Source data are provided (see Data Availability in the main text).}
\label{supfig1}
\end{figure}

\clearpage

\begin{figure}[t]
\centering
\includegraphics[width=\textwidth]{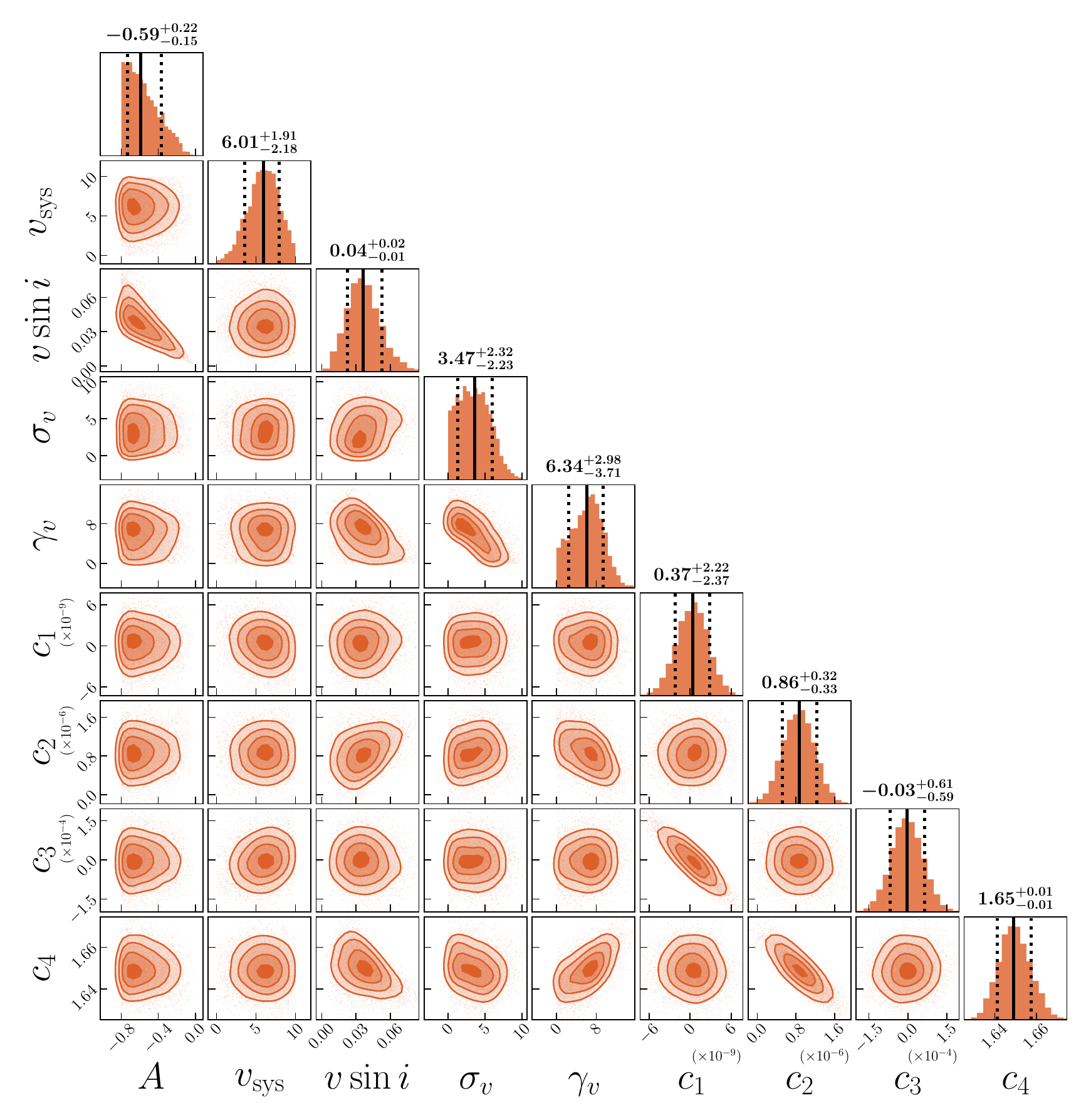}
\caption{\textbf{Posterior of the \pds systemic velocity.} Posterior distributions of the cross-correlated 2018 HARPS spectra with a PHOENIX model spectrum to calculate the \pds systemic velocity. $A$ is the amplitude of the Voigt line, $v_{\rm sys}$ is the systemic velocity in km/s, $v \sin i$ is the projected rotational velocity of the star in km/s (where $i$ is the inclination of the stellar rotation axis), $\sigma_v$ is the Gaussian line width in km/s, $\gamma_v$ is the Lorentzian width in km/s, and $c_1$, $c_2$, $c_3$, and $c_4$ are the polynomial coefficients (see Methods, subsection Derivation of the systemic velocity). Source data are provided (see Data Availability in the main text).}
\label{supfig2}
\end{figure}

\clearpage

\begin{table*}
\centering
\caption{Priors used to fit the Na\,I line components $i$, according to the line (first column), and the continuum. $\mu$ is the centre wavelength of the line in nm, $\sigma_v$ is the Gaussian line width in km/s, $\gamma_v$ is the Lorentzian line width in km/s, $\tau$ is the optical depth, and $f$ is the surface fill fraction. Note that the photospheric absorption line and the stationary emission line have the surface fill fraction fixed to 1. $c_0$ and $c_1$ are the offset and linear polynomial coefficients, respectively, for the first spectral section (around Na\,I D$_2$), while $d_0$ and $d_1$ are the same coefficients for the second section (around Na\,I D$_1$). All priors are assumed to be uniform across the range of values, with the exception of the optical depth, which uses log-uniform priors. This table is also provided in electronic form (see source data in Data Availability in the main text).}
\par\vspace{0.5cm}
\resizebox{\textwidth}{!}{%
\begin{tabular}{lccccc}
\hline
Line component & $\mu$ & $\sigma_v$ & $\gamma_v$ & $\tau$ & $f$ \\ \hline
stellar photospheric absorption & [589.156, 589.160] & [1, 20] & [20, 60] & [0.001, 5] & 1 \\
stationary emission & [589.156, 589.160] & [1, 40] & - & [0.001, 5] & 1 \\
stationary absorption 1 & [589.113, 589.117] & [0.001, 3] & - & [0.001, 5] & [0, 1] \\
stationary absorption 2 & [589.125, 589.129] & [0.001, 3] & - & [0.001, 5] & [0, 1] \\
variable absorption & - & - & - & [0.001, 5] & [0, 1] \\ \hline
& & & & & \\ \hline
Continuum & $c_0$ & $c_1$ & $d_0$ & $d_1$ & - \\ \hline
polynomial coefficient & [1, 2] & [$-0.002$, 0] & [1, 2] & [$-0.002$, 0] & - \\
\hline
\end{tabular}
}
\label{suptable1}
\end{table*}

\clearpage

\begin{table*}
\centering
\caption{$\mu$ and $\sigma_v$ priors used to fit the variable Na\,I absorption line components, where $\mu$ is the centre wavelength of the line in nm, and $\sigma_v$ is the Gaussian line width in km/s. Components correspond to different variable absorption lines present in the same epoch, numbered from 1 (blue-most) to 4 (red-most). This table is also provided in electronic form (see source data in Data Availability in the main text).}
\par\vspace{0.5cm}
\resizebox{\textwidth}{!}{%
\begin{tabular}{lcccccccc}
\hline
\multicolumn{1}{c}{} & \multicolumn{2}{c}{component 1} & \multicolumn{2}{c}{component 2} & \multicolumn{2}{c}{component 3} & \multicolumn{2}{c}{component 4} \\
\cmidrule(lr){2-3} \cmidrule(lr){4-5} \cmidrule(lr){6-7} \cmidrule(lr){8-9}
 Date & $\mu$ & $\sigma_{v}$ & $\mu$ & $\sigma_{v}$ & $\mu$ & $\sigma_{v}$ & $\mu$ & $\sigma_{v}$ \\
\hline
 2018-03-29 (1) & [588.990, 589.000] & [1, 8] & [589.033, 589.043] & [1, 8] & [589.090, 589.100] & [1, 8] & - & - \\
 2018-03-29 (2) & [588.990, 589.000] & [1, 8] & [589.033, 589.043] & [1, 8] & [589.090, 589.100] & [1, 8] & - & - \\
 2018-03-29 (3) & [588.990, 589.000] & [1, 8] & [589.033, 589.043] & [1, 8] & [589.090, 589.100] & [1, 8] & - & - \\
 2018-03-30 (1) & [589.038, 589.042] & [1, 8] & [589.063, 589.073] & [1, 8] & - & - & - & - \\
 2018-03-30 (2) & [589.040, 589.044] & [1, 8] & [589.063, 589.073] & [1, 8] & - & - & - & - \\
 2018-03-30 (3) & [589.043, 589.047] & [1, 8] & [589.063, 589.073] & [1, 8] & - & - & - & - \\
 2018-03-31 (1) & [589.035, 589.045] & [1, 10] & [589.075, 589.085] & [1, 8] & - & - & - & - \\
 2018-03-31 (2) & [589.035, 589.045] & [1, 8] & [589.075, 589.085] & [1, 8] & - & - & - & - \\
 2018-03-31 (3) & [589.030, 589.040] & [1, 10] & [589.075, 589.085] & [1, 8] & - & - & - & - \\
 2018-04-18 & [589.085, 589.105] & [1, 8] & - & - & - & - & - & - \\
 2018-04-19 & [589.095, 589.105] & [1, 3] & - & - & - & - & - & - \\
 2018-04-20 & [589.057, 589.067] & [1, 8] & [589.068, 589.088] & [1, 8] & - & - & - & - \\
 2018-04-21 & [588.928, 588.948] & [1, 8] & [589.036, 589.056] & [1, 5] & [589.068, 589.088] & [1, 3] & [589.098, 589.118] & [1, 2] \\
 2018-04-22 & [589.035, 589.045] & [1, 8] & [589.053, 589.073] & [1, 8] & [589.095, 589.105] & [1, 3] & - & - \\
 2018-04-23 & [589.019, 589.029] & [1, 5] & [589.048, 589.068] & [1, 5] & [589.080, 589.100] & [1, 5] & - & - \\
 2018-05-01 & [589.092, 589.102] & [0.1, 5] & - & - & - & - & - & - \\
 2018-05-06 & [588.945, 588.955] & [1, 5] & [588.993, 589.003] & [1, 5] & [589.029, 589.039] & [1, 8] & [589.095, 589.105] & [1, 15] \\
 2018-05-13 & [589.028, 589.032] & [1, 8] & [589.048, 589.052] & [1, 5] & [589.073, 589.077] & [1, 3] & - & - \\
\hline
\end{tabular}
}
\label{suptable2}
\end{table*}

\clearpage

\begin{table*}
\centering
\caption{Best fit median values found for the stellar photospheric Na\,I absorption line, the stationary emission line, and the partially blended stationary Na\,I absorption lines, where ``1'' stands for the blue-most feature and ``2'' stands for the red-most feature. $v$ is the radial velocity in km/s with respect to the stellar rest frame, $\sigma_v$ is the Gaussian line width in km/s, $\gamma_v$ is the Lorentzian line width in km/s, $\tau$ is the optical depth, $f$ is the surface fill fraction, and $N_{\rm col}$ is the column density in cm$^{-2}$, calculated from the posterior distribution. Uncertainties correspond to the difference between the median value and the 16 and 84 percentiles of the posterior distribution. This table is also provided in electronic form (see source data in Data Availability in the main text).}
\vspace{0.5cm}
\resizebox{\textwidth}{!}{%
\begin{tabular}{lccccccccccccccccccc}
\hline
\multicolumn{1}{c}{} & \multicolumn{5}{c}{photospheric absorption} & \multicolumn{4}{c}{stationary emission} & \multicolumn{5}{c}{stationary absorption 1} & \multicolumn{5}{c}{stationary absorption 2} \\
\cmidrule(lr){2-6} \cmidrule(lr){7-10} \cmidrule(lr){11-15} \cmidrule(lr){16-20}
 Date & $v$ & $\sigma_{v}$ & $\gamma_{v}$ & $\tau$ & $N_{\rm col}\,(\times10^{12})$ & $v$ & $\sigma_{v}$ & $\tau$ & $N_{\rm col}\,(\times10^{12})$ & $v$ & $\sigma_{v}$ & $\tau$ & $f$ & $N_{\rm col}\,(\times10^{12})$ & $v$ & $\sigma_{v}$ & $\tau$ & $f$ & $N_{\rm col}\,(\times10^{12})$ \\
\hline
 2018-03-29 (1) & $0.83^{+0.14}_{-0.30}$ & 16.65$^{+1.25}_{-1.30}$ & 24.45$^{+1.25}_{-1.25}$ & 3.18$^{+0.05}_{-0.05}$ & 32.39$^{+0.54}_{-0.53}$ & $0.79^{+0.17}_{-0.37}$ & 39.90$^{+0.08}_{-0.17}$ & 0.77$^{+0.01}_{-0.01}$ & 7.68$^{+0.08}_{-0.08}$ & $-21.09^{+0.05}_{-0.09}$ & 2.58$^{+0.18}_{-0.17}$ & 0.82$^{+0.13}_{-0.08}$ & 0.93$^{+0.05}_{-0.09}$ & 0.53$^{+0.09}_{-0.06}$ & $-14.98^{+0.04}_{-0.06}$ & 2.99$^{+0.01}_{-0.02}$ & 2.91$^{+0.13}_{-0.13}$ & 0.86$^{+0.01}_{-0.01}$ & 2.17$^{+0.09}_{-0.09}$ \\
 2018-03-29 (2) & $0.96^{+0.04}_{-0.09}$ & 17.24$^{+1.19}_{-1.19}$ & 23.66$^{+1.15}_{-1.18}$ & 3.11$^{+0.05}_{-0.05}$ & 31.43$^{+0.50}_{-0.50}$ & $-0.24^{+0.74}_{-0.55}$ & 39.90$^{+0.08}_{-0.17}$ & 0.78$^{+0.01}_{-0.01}$ & 7.82$^{+0.08}_{-0.08}$ & $-21.04^{+0.01}_{-0.02}$ & 2.13$^{+0.13}_{-0.13}$ & 1.39$^{+0.44}_{-0.37}$ & 0.65$^{+0.13}_{-0.09}$ & 0.74$^{+0.20}_{-0.18}$ & $-15.33^{+0.06}_{-0.06}$ & 3.00$^{+0.00}_{-0.01}$ & 2.50$^{+0.11}_{-0.11}$ & 0.91$^{+0.02}_{-0.01}$ & 1.87$^{+0.08}_{-0.08}$ \\
 2018-03-29 (3) & $0.98^{+0.03}_{-0.07}$ & 19.13$^{+0.38}_{-0.47}$ & 20.37$^{+0.50}_{-0.27}$ & 3.00$^{+0.04}_{-0.04}$ & 29.24$^{+0.35}_{-0.34}$ & $-0.87^{+0.23}_{-0.11}$ & 39.86$^{+0.10}_{-0.23}$ & 0.79$^{+0.01}_{-0.01}$ & 7.84$^{+0.07}_{-0.07}$ & $-21.55^{+0.10}_{-0.09}$ & 1.43$^{+0.12}_{-0.10}$ & 0.97$^{+0.10}_{-0.07}$ & 0.95$^{+0.03}_{-0.06}$ & 0.35$^{+0.05}_{-0.04}$ & $-15.48^{+0.07}_{-0.07}$ & 3.00$^{+0.00}_{-0.01}$ & 2.92$^{+0.12}_{-0.12}$ & 0.89$^{+0.01}_{-0.01}$ & 2.18$^{+0.09}_{-0.09}$ \\
 2018-03-30 (1) & $0.99^{+0.02}_{-0.05}$ & 19.81$^{+0.14}_{-0.31}$ & 32.78$^{+0.77}_{-0.81}$ & 2.83$^{+0.05}_{-0.05}$ & 36.90$^{+0.58}_{-0.63}$ & $-0.51^{+0.42}_{-0.32}$ & 39.58$^{+0.31}_{-0.65}$ & 0.85$^{+0.01}_{-0.01}$ & 8.36$^{+0.13}_{-0.15}$ & $-21.12^{+0.06}_{-0.09}$ & 2.70$^{+0.11}_{-0.11}$ & 1.68$^{+0.25}_{-0.23}$ & 0.83$^{+0.06}_{-0.05}$ & 1.14$^{+0.15}_{-0.15}$ & $-14.93^{+0.00}_{-0.01}$ & 3.00$^{+0.00}_{-0.00}$ & 2.96$^{+0.14}_{-0.14}$ & 0.90$^{+0.01}_{-0.01}$ & 2.22$^{+0.11}_{-0.10}$ \\
 2018-03-30 (2) & $1.00^{+0.01}_{-0.03}$ & 19.80$^{+0.15}_{-0.33}$ & 30.14$^{+1.37}_{-1.12}$ & 2.94$^{+0.06}_{-0.06}$ & 36.27$^{+1.32}_{-1.06}$ & $0.51^{+0.38}_{-0.76}$ & 37.09$^{+1.71}_{-1.11}$ & 0.89$^{+0.01}_{-0.01}$ & 8.23$^{+0.34}_{-0.25}$ & $-21.13^{+0.06}_{-0.09}$ & 2.25$^{+0.11}_{-0.11}$ & 1.47$^{+0.22}_{-0.16}$ & 0.91$^{+0.06}_{-0.06}$ & 0.83$^{+0.11}_{-0.08}$ & $-14.93^{+0.01}_{-0.02}$ & 3.00$^{+0.00}_{-0.01}$ & 3.01$^{+0.17}_{-0.16}$ & 0.86$^{+0.02}_{-0.02}$ & 2.26$^{+0.12}_{-0.12}$ \\
 2018-03-30 (3) & $1.01^{+0.01}_{-0.02}$ & 19.78$^{+0.16}_{-0.36}$ & 29.62$^{+0.82}_{-0.81}$ & 2.69$^{+0.06}_{-0.05}$ & 32.77$^{+0.72}_{-0.70}$ & $0.86^{+0.12}_{-0.25}$ & 33.04$^{+0.63}_{-0.57}$ & 0.90$^{+0.01}_{-0.01}$ & 7.46$^{+0.16}_{-0.16}$ & $-21.36^{+0.10}_{-0.10}$ & 2.14$^{+0.10}_{-0.10}$ & 1.31$^{+0.12}_{-0.07}$ & 0.97$^{+0.03}_{-0.04}$ & 0.71$^{+0.06}_{-0.05}$ & $-14.93^{+0.01}_{-0.02}$ & 3.00$^{+0.00}_{-0.01}$ & 3.68$^{+0.19}_{-0.18}$ & 0.83$^{+0.01}_{-0.01}$ & 2.75$^{+0.15}_{-0.13}$ \\
 2018-03-31 (1) & $-0.98^{+0.07}_{-0.03}$ & 19.82$^{+0.14}_{-0.29}$ & 24.94$^{+0.43}_{-0.38}$ & 3.28$^{+0.04}_{-0.03}$ & 36.21$^{+0.32}_{-0.32}$ & $0.81^{+0.15}_{-0.25}$ & 39.91$^{+0.07}_{-0.15}$ & 0.75$^{+0.01}_{-0.01}$ & 7.51$^{+0.05}_{-0.06}$ & $-21.67^{+0.07}_{-0.07}$ & 2.05$^{+0.07}_{-0.07}$ & 1.06$^{+0.11}_{-0.06}$ & 0.95$^{+0.03}_{-0.06}$ & 0.54$^{+0.05}_{-0.03}$ & $-14.94^{+0.01}_{-0.02}$ & 3.00$^{+0.00}_{-0.01}$ & 2.64$^{+0.11}_{-0.11}$ & 0.86$^{+0.01}_{-0.01}$ & 1.98$^{+0.08}_{-0.08}$ \\
 2018-03-31 (2) & $-0.97^{+0.07}_{-0.03}$ & 19.83$^{+0.13}_{-0.27}$ & 27.23$^{+0.42}_{-0.38}$ & 3.22$^{+0.03}_{-0.03}$ & 37.34$^{+0.30}_{-0.30}$ & $0.90^{+0.09}_{-0.17}$ & 39.95$^{+0.04}_{-0.09}$ & 0.75$^{+0.00}_{-0.01}$ & 7.50$^{+0.05}_{-0.05}$ & $-21.48^{+0.10}_{-0.11}$ & 2.28$^{+0.09}_{-0.10}$ & 0.98$^{+0.05}_{-0.03}$ & 0.98$^{+0.02}_{-0.03}$ & 0.56$^{+0.03}_{-0.03}$ & $-14.97^{+0.04}_{-0.06}$ & 3.00$^{+0.00}_{-0.01}$ & 2.12$^{+0.10}_{-0.09}$ & 0.92$^{+0.02}_{-0.02}$ & 1.59$^{+0.07}_{-0.07}$ \\
 2018-03-31 (3) & $-0.94^{+0.11}_{-0.06}$ & 19.88$^{+0.09}_{-0.20}$ & 25.47$^{+0.36}_{-0.35}$ & 3.18$^{+0.03}_{-0.03}$ & 35.55$^{+0.29}_{-0.29}$ & $0.94^{+0.06}_{-0.12}$ & 39.93$^{+0.05}_{-0.11}$ & 0.74$^{+0.01}_{-0.01}$ & 7.41$^{+0.05}_{-0.05}$ & $-21.90^{+0.09}_{-0.08}$ & 2.04$^{+0.10}_{-0.09}$ & 0.98$^{+0.04}_{-0.03}$ & 0.98$^{+0.01}_{-0.02}$ & 0.50$^{+0.03}_{-0.02}$ & $-15.07^{+0.06}_{-0.06}$ & 3.00$^{+0.00}_{-0.01}$ & 2.14$^{+0.08}_{-0.08}$ & 0.96$^{+0.01}_{-0.01}$ & 1.60$^{+0.06}_{-0.06}$ \\
 2018-04-18 & $-0.47^{+0.18}_{-0.18}$ & 8.71$^{+1.40}_{-1.70}$ & 29.19$^{+0.89}_{-0.86}$ & 3.83$^{+0.05}_{-0.05}$ & 37.55$^{+0.37}_{-0.38}$ & $0.92^{+0.07}_{-0.16}$ & 39.92$^{+0.06}_{-0.14}$ & 0.61$^{+0.00}_{-0.00}$ & 6.12$^{+0.05}_{-0.05}$ & $-22.54^{+0.11}_{-0.11}$ & 1.46$^{+0.12}_{-0.11}$ & 0.72$^{+0.14}_{-0.08}$ & 0.91$^{+0.07}_{-0.11}$ & 0.27$^{+0.05}_{-0.04}$ & $-15.38^{+0.07}_{-0.07}$ & 2.97$^{+0.02}_{-0.04}$ & 1.71$^{+0.09}_{-0.06}$ & 0.98$^{+0.02}_{-0.02}$ & 1.27$^{+0.06}_{-0.04}$ \\
 2018-04-19 & $-0.01^{+0.21}_{-0.22}$ & 19.56$^{+0.31}_{-0.57}$ & 21.15$^{+0.75}_{-0.61}$ & 3.73$^{+0.07}_{-0.06}$ & 37.38$^{+0.55}_{-0.56}$ & $0.93^{+0.07}_{-0.14}$ & 39.51$^{+0.36}_{-0.68}$ & 0.68$^{+0.01}_{-0.01}$ & 6.68$^{+0.10}_{-0.12}$ & $-22.52^{+0.10}_{-0.10}$ & 1.01$^{+0.10}_{-0.09}$ & 2.56$^{+1.04}_{-0.82}$ & 0.59$^{+0.10}_{-0.05}$ & 0.64$^{+0.19}_{-0.17}$ & $-15.50^{+0.07}_{-0.07}$ & 2.35$^{+0.08}_{-0.08}$ & 1.73$^{+0.08}_{-0.06}$ & 0.99$^{+0.01}_{-0.01}$ & 1.02$^{+0.04}_{-0.03}$ \\
 2018-04-20 & $0.91^{+0.07}_{-0.11}$ & 19.90$^{+0.08}_{-0.16}$ & 20.11$^{+0.16}_{-0.08}$ & 2.70$^{+0.03}_{-0.03}$ & 26.54$^{+0.29}_{-0.30}$ & $0.99^{+0.02}_{-0.04}$ & 34.15$^{+0.30}_{-0.32}$ & 0.87$^{+0.00}_{-0.00}$ & 7.47$^{+0.09}_{-0.09}$ & $-23.00^{+0.05}_{-0.04}$ & 1.87$^{+0.08}_{-0.07}$ & 0.80$^{+0.11}_{-0.06}$ & 0.93$^{+0.05}_{-0.08}$ & 0.37$^{+0.05}_{-0.03}$ & $-15.38^{+0.04}_{-0.04}$ & 2.98$^{+0.01}_{-0.02}$ & 1.59$^{+0.03}_{-0.02}$ & 0.99$^{+0.00}_{-0.01}$ & 1.18$^{+0.02}_{-0.02}$ \\
 2018-04-21 & $-1.00^{+0.02}_{-0.01}$ & 11.48$^{+1.23}_{-1.37}$ & 29.81$^{+0.99}_{-0.99}$ & 3.72$^{+0.05}_{-0.05}$ & 39.03$^{+0.44}_{-0.46}$ & $0.99^{+0.02}_{-0.04}$ & 39.94$^{+0.04}_{-0.09}$ & 0.67$^{+0.01}_{-0.01}$ & 6.66$^{+0.05}_{-0.05}$ & $-22.55^{+0.16}_{-0.25}$ & 1.27$^{+0.16}_{-0.11}$ & 3.89$^{+0.55}_{-0.49}$ & 0.66$^{+0.03}_{-0.03}$ & 1.24$^{+0.16}_{-0.14}$ & $-15.90^{+0.08}_{-0.08}$ & 2.94$^{+0.04}_{-0.06}$ & 1.54$^{+0.05}_{-0.04}$ & 0.99$^{+0.01}_{-0.01}$ & 1.13$^{+0.03}_{-0.03}$ \\
 2018-04-22 & $-0.09^{+0.24}_{-0.23}$ & 19.07$^{+0.64}_{-0.99}$ & 25.10$^{+1.03}_{-0.73}$ & 3.29$^{+0.04}_{-0.04}$ & 35.85$^{+0.42}_{-0.40}$ & $0.89^{+0.10}_{-0.21}$ & 39.92$^{+0.06}_{-0.12}$ & 0.71$^{+0.01}_{-0.01}$ & 7.12$^{+0.06}_{-0.06}$ & $-22.50^{+0.12}_{-0.12}$ & 1.62$^{+0.11}_{-0.10}$ & 1.28$^{+0.30}_{-0.21}$ & 0.84$^{+0.10}_{-0.09}$ & 0.52$^{+0.11}_{-0.08}$ & $-16.10^{+0.08}_{-0.08}$ & 2.71$^{+0.08}_{-0.08}$ & 1.68$^{+0.11}_{-0.08}$ & 0.97$^{+0.02}_{-0.03}$ & 1.14$^{+0.07}_{-0.05}$ \\
 2018-04-23 & $0.94^{+0.06}_{-0.11}$ & 19.18$^{+0.57}_{-0.84}$ & 23.55$^{+0.91}_{-0.69}$ & 3.25$^{+0.04}_{-0.04}$ & 34.21$^{+0.46}_{-0.44}$ & $0.93^{+0.06}_{-0.13}$ & 39.83$^{+0.13}_{-0.27}$ & 0.77$^{+0.01}_{-0.01}$ & 7.66$^{+0.07}_{-0.08}$ & $-23.02^{+0.06}_{-0.03}$ & 2.17$^{+0.13}_{-0.12}$ & 1.34$^{+0.31}_{-0.28}$ & 0.73$^{+0.11}_{-0.08}$ & 0.73$^{+0.14}_{-0.13}$ & $-15.95^{+0.07}_{-0.07}$ & 2.89$^{+0.07}_{-0.08}$ & 1.72$^{+0.12}_{-0.09}$ & 0.96$^{+0.02}_{-0.03}$ & 1.24$^{+0.07}_{-0.06}$ \\
 2018-05-01 & $0.37^{+0.21}_{-0.27}$ & 6.80$^{+2.06}_{-2.77}$ & 34.26$^{+0.97}_{-1.05}$ & 3.54$^{+0.05}_{-0.05}$ & 38.91$^{+0.42}_{-0.44}$ & $0.67^{+0.25}_{-0.47}$ & 39.92$^{+0.06}_{-0.13}$ & 0.72$^{+0.01}_{-0.01}$ & 7.23$^{+0.06}_{-0.06}$ & $-22.20^{+0.12}_{-0.13}$ & 1.19$^{+0.10}_{-0.09}$ & 3.95$^{+0.67}_{-0.71}$ & 0.58$^{+0.05}_{-0.04}$ & 1.17$^{+0.17}_{-0.19}$ & $-16.90^{+0.07}_{-0.04}$ & 2.92$^{+0.05}_{-0.07}$ & 2.06$^{+0.16}_{-0.15}$ & 0.93$^{+0.03}_{-0.03}$ & 1.51$^{+0.11}_{-0.10}$ \\
 2018-05-06 & $0.93^{+0.06}_{-0.10}$ & 12.70$^{+0.84}_{-0.89}$ & 29.14$^{+0.76}_{-0.78}$ & 3.38$^{+0.03}_{-0.03}$ & 35.66$^{+0.34}_{-0.35}$ & $-0.96^{+0.09}_{-0.04}$ & 39.88$^{+0.09}_{-0.19}$ & 0.86$^{+0.00}_{-0.00}$ & 8.54$^{+0.05}_{-0.05}$ & $-22.74^{+0.08}_{-0.08}$ & 1.66$^{+0.09}_{-0.09}$ & 0.79$^{+0.24}_{-0.16}$ & 0.79$^{+0.14}_{-0.13}$ & 0.33$^{+0.09}_{-0.06}$ & $-16.07^{+0.05}_{-0.05}$ & 2.47$^{+0.06}_{-0.06}$ & 1.13$^{+0.03}_{-0.02}$ & 0.99$^{+0.01}_{-0.01}$ & 0.70$^{+0.02}_{-0.02}$ \\
 2018-05-13 & $-0.41^{+0.21}_{-0.23}$ & 19.92$^{+0.06}_{-0.13}$ & 20.12$^{+0.18}_{-0.09}$ & 3.13$^{+0.03}_{-0.03}$ & 30.89$^{+0.34}_{-0.33}$ & $0.79^{+0.16}_{-0.27}$ & 34.18$^{+0.41}_{-0.40}$ & 0.84$^{+0.00}_{-0.01}$ & 7.15$^{+0.10}_{-0.10}$ & $-22.39^{+0.07}_{-0.07}$ & 1.74$^{+0.09}_{-0.08}$ & 1.20$^{+0.27}_{-0.25}$ & 0.73$^{+0.11}_{-0.08}$ & 0.52$^{+0.10}_{-0.09}$ & $-15.81^{+0.05}_{-0.05}$ & 2.24$^{+0.06}_{-0.06}$ & 1.22$^{+0.03}_{-0.03}$ & 0.99$^{+0.01}_{-0.01}$ & 0.69$^{+0.02}_{-0.02}$ \\
\hline
\end{tabular}
}
\label{suptable3}
\end{table*}

\clearpage

\begin{table*}
\centering
\caption{Best fit median values found for the variable Na\,I absorption lines. $v$ is the radial velocity in km/s with respect to the stellar rest frame, $\sigma_v$ is the Gaussian line width in km/s, $\tau$ is the optical depth, $f$ is the surface fill fraction, and $N_{\rm col}$ is the column density in cm$^{-2}$, calculated from the posterior distribution. Uncertainties correspond to the difference between the median value and the 16 and 84 percentiles of the posterior distribution. This table is also provided in electronic form (see source data in Data Availability in the main text).}
\vspace{0.5cm}
\resizebox{\textwidth}{!}{%
\begin{tabular}{lcccccccccccccccccccc}
\hline
\multicolumn{1}{c}{} & \multicolumn{5}{c}{component 1} & \multicolumn{5}{c}{component 2} & \multicolumn{5}{c}{component 3} & \multicolumn{5}{c}{component 4} \\
\cmidrule(lr){2-6} \cmidrule(lr){7-11} \cmidrule(lr){12-16} \cmidrule(lr){17-21}
 Date & $v$ & $\sigma_{v}$ & $\tau$ & $f$ & $N_{\rm col}\,(\times10^{12})$ & $v$ & $\sigma_{v}$ & $\tau$ & $f$ & $N_{\rm col}\,(\times10^{12})$ & $v$ & $\sigma_{v}$ & $\tau$ & $f$ & $N_{\rm col}\,(\times10^{12})$ & $v$ & $\sigma_{v}$ & $\tau$ & $f$ & $N_{\rm col}\,(\times10^{12})$ \\
\hline
 2018-03-29 (1) & $-80.93^{+0.23}_{-0.28}$ & 6.08$^{+0.17}_{-0.17}$ & 3.83$^{+0.37}_{-0.33}$ & 0.40$^{+0.01}_{-0.01}$ & 5.83$^{+0.49}_{-0.45}$ & $-59.50^{+0.16}_{-0.18}$ & 6.05$^{+0.15}_{-0.14}$ & 2.57$^{+0.15}_{-0.14}$ & 0.65$^{+0.01}_{-0.01}$ & 3.90$^{+0.20}_{-0.20}$ & $-29.96^{+0.20}_{-0.33}$ & 5.43$^{+0.32}_{-0.35}$ & 1.51$^{+0.22}_{-0.21}$ & 0.59$^{+0.05}_{-0.04}$ & 2.05$^{+0.28}_{-0.29}$ & - & - & - & - & - \\
 2018-03-29 (2) & $-80.91^{+0.21}_{-0.24}$ & 6.21$^{+0.16}_{-0.16}$ & 2.60$^{+0.20}_{-0.19}$ & 0.47$^{+0.01}_{-0.01}$ & 4.03$^{+0.28}_{-0.28}$ & $-59.40^{+0.15}_{-0.17}$ & 6.33$^{+0.16}_{-0.14}$ & 2.44$^{+0.13}_{-0.13}$ & 0.66$^{+0.01}_{-0.01}$ & 3.87$^{+0.18}_{-0.17}$ & $-29.81^{+0.10}_{-0.21}$ & 5.68$^{+0.28}_{-0.32}$ & 1.96$^{+0.23}_{-0.21}$ & 0.51$^{+0.03}_{-0.02}$ & 2.78$^{+0.32}_{-0.30}$ & - & - & - & - & - \\
 2018-03-29 (3) & $-80.78^{+0.14}_{-0.18}$ & 5.75$^{+0.12}_{-0.13}$ & 2.75$^{+0.17}_{-0.16}$ & 0.53$^{+0.01}_{-0.01}$ & 3.95$^{+0.20}_{-0.20}$ & $-59.60^{+0.12}_{-0.14}$ & 6.66$^{+0.14}_{-0.13}$ & 1.95$^{+0.11}_{-0.10}$ & 0.73$^{+0.02}_{-0.02}$ & 3.25$^{+0.15}_{-0.15}$ & $-30.55^{+0.30}_{-0.27}$ & 6.00$^{+0.30}_{-0.28}$ & 1.55$^{+0.17}_{-0.17}$ & 0.59$^{+0.04}_{-0.03}$ & 2.33$^{+0.24}_{-0.24}$ & - & - & - & - & - \\
 2018-03-30 (1) & $-60.81^{+0.26}_{-0.23}$ & 2.83$^{+0.12}_{-0.11}$ & 4.62$^{+0.28}_{-0.49}$ & 0.30$^{+0.02}_{-0.01}$ & 3.24$^{+0.19}_{-0.28}$ & $-46.76^{+0.17}_{-0.16}$ & 4.99$^{+0.11}_{-0.12}$ & 4.96$^{+0.03}_{-0.06}$ & 0.62$^{+0.01}_{-0.01}$ & 6.17$^{+0.15}_{-0.16}$ & - & - & - & - & - & - & - & - & - & - \\
 2018-03-30 (2) & $-59.03^{+0.37}_{-0.35}$ & 3.09$^{+0.21}_{-0.17}$ & 3.41$^{+0.98}_{-0.89}$ & 0.31$^{+0.04}_{-0.02}$ & 2.64$^{+0.60}_{-0.61}$ & $-45.40^{+0.13}_{-0.13}$ & 3.99$^{+0.09}_{-0.08}$ & 4.97$^{+0.02}_{-0.05}$ & 0.71$^{+0.01}_{-0.01}$ & 4.94$^{+0.12}_{-0.11}$ & - & - & - & - & - & - & - & - & - & - \\
 2018-03-30 (3) & $-57.32^{+0.29}_{-0.30}$ & 3.25$^{+0.37}_{-0.38}$ & 0.72$^{+0.56}_{-0.27}$ & 0.57$^{+0.25}_{-0.18}$ & 0.59$^{+0.37}_{-0.21}$ & $-44.33^{+0.12}_{-0.11}$ & 4.06$^{+0.07}_{-0.07}$ & 4.98$^{+0.01}_{-0.03}$ & 0.73$^{+0.01}_{-0.01}$ & 5.04$^{+0.09}_{-0.09}$ & - & - & - & - & - & - & - & - & - & - \\
 2018-03-31 (1) & $-57.69^{+0.02}_{-0.04}$ & 9.92$^{+0.06}_{-0.10}$ & 2.86$^{+0.12}_{-0.11}$ & 0.54$^{+0.01}_{-0.01}$ & 7.07$^{+0.28}_{-0.26}$ & $-39.40^{+0.07}_{-0.07}$ & 4.28$^{+0.05}_{-0.05}$ & 4.97$^{+0.03}_{-0.05}$ & 0.71$^{+0.00}_{-0.00}$ & 5.30$^{+0.06}_{-0.07}$ & - & - & - & - & - & - & - & - & - & - \\
 2018-03-31 (2) & $-60.80^{+0.16}_{-0.16}$ & 6.62$^{+0.14}_{-0.14}$ & 2.63$^{+0.11}_{-0.10}$ & 0.69$^{+0.01}_{-0.01}$ & 4.35$^{+0.14}_{-0.14}$ & $-41.25^{+0.12}_{-0.12}$ & 4.84$^{+0.06}_{-0.06}$ & 4.96$^{+0.03}_{-0.06}$ & 0.70$^{+0.00}_{-0.00}$ & 5.99$^{+0.08}_{-0.09}$ & - & - & - & - & - & - & - & - & - & - \\
 2018-03-31 (3) & $-61.44^{+0.16}_{-0.15}$ & 7.08$^{+0.12}_{-0.12}$ & 2.63$^{+0.09}_{-0.08}$ & 0.72$^{+0.01}_{-0.01}$ & 4.65$^{+0.14}_{-0.14}$ & $-41.75^{+0.11}_{-0.11}$ & 4.77$^{+0.08}_{-0.08}$ & 4.56$^{+0.17}_{-0.17}$ & 0.70$^{+0.01}_{-0.01}$ & 5.45$^{+0.16}_{-0.16}$ & - & - & - & - & - & - & - & - & - & - \\
 2018-04-18 & $-31.64^{+0.19}_{-0.19}$ & 5.31$^{+0.17}_{-0.16}$ & 2.26$^{+0.18}_{-0.17}$ & 0.68$^{+0.02}_{-0.02}$ & 2.99$^{+0.22}_{-0.21}$ & - & - & - & - & - & - & - & - & - & - & - & - & - & - & - \\
 2018-04-19 & $-31.83^{+0.17}_{-0.16}$ & 2.99$^{+0.01}_{-0.01}$ & 4.91$^{+0.06}_{-0.14}$ & 0.45$^{+0.01}_{-0.01}$ & 3.67$^{+0.05}_{-0.11}$ & - & - & - & - & - & - & - & - & - & - & - & - & - & - & - \\
 2018-04-20 & $-48.34^{+0.18}_{-0.18}$ & 2.57$^{+0.12}_{-0.11}$ & 3.43$^{+0.51}_{-0.42}$ & 0.32$^{+0.02}_{-0.02}$ & 2.21$^{+0.26}_{-0.22}$ & $-36.74^{+0.10}_{-0.10}$ & 4.46$^{+0.11}_{-0.10}$ & 2.58$^{+0.11}_{-0.11}$ & 0.70$^{+0.01}_{-0.01}$ & 2.88$^{+0.10}_{-0.10}$ & - & - & - & - & - & - & - & - & - & - \\
 2018-04-21 & $-111.59^{+0.20}_{-0.19}$ & 6.98$^{+0.27}_{-0.28}$ & 0.57$^{+0.23}_{-0.14}$ & 0.70$^{+0.19}_{-0.16}$ & 1.00$^{+0.35}_{-0.24}$ & $-53.84^{+0.16}_{-0.16}$ & 4.99$^{+0.01}_{-0.02}$ & 1.98$^{+0.20}_{-0.19}$ & 0.51$^{+0.03}_{-0.02}$ & 2.47$^{+0.25}_{-0.24}$ & $-38.14^{+0.09}_{-0.09}$ & 2.99$^{+0.00}_{-0.01}$ & 4.97$^{+0.03}_{-0.05}$ & 0.75$^{+0.01}_{-0.01}$ & 3.71$^{+0.02}_{-0.04}$ & $-27.26^{+0.16}_{-0.25}$ & 1.99$^{+0.01}_{-0.02}$ & 4.92$^{+0.06}_{-0.14}$ & 0.70$^{+0.01}_{-0.02}$ & 2.44$^{+0.04}_{-0.07}$ \\
 2018-04-22 & $-58.63^{+0.43}_{-0.37}$ & 3.92$^{+0.46}_{-0.42}$ & 0.74$^{+0.75}_{-0.40}$ & 0.38$^{+0.33}_{-0.14}$ & 0.74$^{+0.61}_{-0.37}$ & $-44.73^{+0.08}_{-0.08}$ & 4.02$^{+0.05}_{-0.05}$ & 4.88$^{+0.08}_{-0.13}$ & 0.81$^{+0.00}_{-0.00}$ & 4.89$^{+0.09}_{-0.10}$ & $-28.14^{+0.14}_{-0.14}$ & 2.99$^{+0.01}_{-0.02}$ & 3.49$^{+0.23}_{-0.22}$ & 0.69$^{+0.01}_{-0.01}$ & 2.60$^{+0.17}_{-0.16}$ & - & - & - & - & - \\
 2018-04-23 & $-65.85^{+0.03}_{-0.07}$ & 4.59$^{+0.25}_{-0.29}$ & 3.87$^{+0.75}_{-0.96}$ & 0.11$^{+0.01}_{-0.01}$ & 4.39$^{+0.82}_{-1.04}$ & $-48.34^{+0.06}_{-0.06}$ & 2.72$^{+0.03}_{-0.04}$ & 4.96$^{+0.03}_{-0.07}$ & 0.85$^{+0.00}_{-0.00}$ & 3.37$^{+0.05}_{-0.05}$ & $-37.07^{+0.31}_{-0.35}$ & 4.82$^{+0.13}_{-0.24}$ & 1.12$^{+0.34}_{-0.31}$ & 0.52$^{+0.13}_{-0.08}$ & 1.33$^{+0.41}_{-0.36}$ & - & - & - & - & - \\
 2018-05-01 & $-28.88^{+0.14}_{-0.18}$ & 3.28$^{+0.10}_{-0.11}$ & 4.44$^{+0.28}_{-0.26}$ & 0.76$^{+0.01}_{-0.01}$ & 3.63$^{+0.22}_{-0.22}$ & - & - & - & - & - & - & - & - & - & - & - & - & - & - & - \\
 2018-05-06 & $-106.05^{+0.16}_{-0.16}$ & 4.97$^{+0.02}_{-0.04}$ & 1.25$^{+0.23}_{-0.22}$ & 0.29$^{+0.04}_{-0.03}$ & 1.56$^{+0.28}_{-0.27}$ & $-82.25^{+0.37}_{-0.38}$ & 4.97$^{+0.02}_{-0.04}$ & 0.23$^{+0.13}_{-0.05}$ & 0.72$^{+0.20}_{-0.24}$ & 0.28$^{+0.16}_{-0.06}$ & $-65.01^{+0.58}_{-0.52}$ & 5.70$^{+0.63}_{-0.59}$ & 0.17$^{+0.21}_{-0.06}$ & 0.57$^{+0.29}_{-0.30}$ & 0.24$^{+0.29}_{-0.08}$ & $-27.20^{+0.05}_{-0.10}$ & 7.37$^{+0.20}_{-0.20}$ & 0.55$^{+0.12}_{-0.07}$ & 0.84$^{+0.10}_{-0.12}$ & 1.01$^{+0.22}_{-0.13}$ \\
 2018-05-13 & $-66.30^{+0.02}_{-0.01}$ & 7.03$^{+0.08}_{-0.08}$ & 2.69$^{+0.10}_{-0.10}$ & 0.72$^{+0.01}_{-0.01}$ & 4.73$^{+0.15}_{-0.14}$ & $-54.11^{+0.00}_{-0.01}$ & 3.02$^{+0.07}_{-0.07}$ & 3.75$^{+0.30}_{-0.27}$ & 0.59$^{+0.01}_{-0.01}$ & 2.84$^{+0.18}_{-0.17}$ & $-43.40^{+0.02}_{-0.01}$ & 3.00$^{+0.00}_{-0.01}$ & 4.89$^{+0.08}_{-0.16}$ & 0.46$^{+0.01}_{-0.01}$ & 3.66$^{+0.06}_{-0.12}$ & - & - & - & - & - \\
\hline
\end{tabular}
}
\label{suptable4}
\end{table*}

\clearpage

\begin{table*}
\centering
\caption{Input orbital parameters used in N-body simulations with \texttt{REBOUND} with planets b and c (2-planet simulations) and b, c, and d (3-planet simulations). Parameters include mass ($M$) in M$_{\rm Jup}$, semi-major axis ($a$) in au, eccentricity ($e$), inclination relative to the disc inclination ($i$) in degrees, argument of periastron ($\omega$) in degrees, and longitude of the ascending node ($\Omega$) in degrees, and stellar mass ($M_\star$) in M$_\odot$. Values in this table correspond to the median value of the posterior of the ``stable (including N-body)'' orbital parameters from Tables 3 and 4 of \citep{Trevascus+25} for the first (2-planet) and second (3-planet), respectively. This table is also provided in electronic form (see source data in Data Availability in the main text).}
\vspace{0.5cm}
\resizebox{\textwidth}{!}{%
\begin{tabular}{llcccccc}
\hline
2-planet simulations & Object & $M$ (M$_{\rm Jup}$) & $a$ (au) & $e$ & $i$ (deg) & $\omega$ (deg) & $\Omega$ (deg) \\ \hline
 & PDS 70 b & 1.4 & 20.7 & 0.160 & 2.3 & 190.0 & 176.0 \\
 & PDS 70 c & 6.4 & 33.9 & 0.042 & 1.5 & 77.0 & 158.0 \\
 & PDS 70 & \multicolumn{6}{c}{$M_\star$ = 0.952 M$_\odot$} \\ \hline
 &  & \multicolumn{1}{l}{} & \multicolumn{1}{l}{} & \multicolumn{1}{l}{} & \multicolumn{1}{l}{} & \multicolumn{1}{l}{} & \multicolumn{1}{l}{} \\ \hline
3-planet simulations & Object & $M$ (M$_{\rm Jup}$) & $a$ (au) & $e$ & $i$ (deg) & $\omega$ (deg) & $\Omega$ (deg) \\ \hline
 & PDS 70 b & 0.7 & 21.1 & 0.131 & 0.4 & 191.4 & 174.3 \\
 & PDS 70 c & 2.4 & 35.3 & 0.033 & 0.2 & 63.0 & 159.8 \\
 & PDS 70 d & 0.4 & 10.7 & 0.250 & 22.7 & 29.0 & 144.0 \\
 & PDS 70 & \multicolumn{6}{c}{$M_\star$ = 0.965 M$_\odot$} \\ \hline
\end{tabular}
}
\label{suptable5}
\end{table*}

\end{document}